\documentclass[ALICE,manyauthors]{cernphprep}
\usepackage[comma,square,numbers,sort&compress]{natbib}
\usepackage{amsmath,amssymb,amsfonts}

\usepackage{xspace}
\usepackage[utf8]{inputenc}
\usepackage[normalem]{ulem}
\usepackage[T1]{fontenc}
\usepackage{units}
\usepackage{bm}

\usepackage{multirow}
\usepackage{dcolumn}
\usepackage{tabularx}
\usepackage{booktabs}
\newcolumntype{C}{>{\centering\arraybackslash}X}

\usepackage{graphics}
\usepackage{graphicx}
\usepackage{subfigure}
\usepackage{grffile}
\usepackage{epsfig}
\usepackage{rotating}
\graphicspath{{figures/}}

\usepackage[dvipsnames]{xcolor}
\usepackage[colorlinks = true,      %
            linkcolor  = RubineRed, %
            urlcolor   = RoyalBlue, %
            citecolor  = Magenta]{hyperref}

\usepackage{listings}
\usepackage{lineno}

\DeclareUnicodeCharacter{00A0}{}

\newcommand{\pp}    {pp\xspace}

\newcommand{\PbPb}  {\mbox{Pb--Pb}\xspace}

\newcommand{\pPb}   {\mbox{p--Pb}\xspace}

\newcommand{\pt}           {\ensuremath{p_{\rm T}}\xspace}

\newcommand{\ycms}         {\ensuremath{y_{\rm CMS}}\xspace}

\newcommand{\dEdx}         {\ensuremath{\textrm{d}E/\textrm{d}x}\xspace}

\newcommand{\nineH}         {$\sqrt{s}~=~0.9$~Te\kern-.1emV\xspace}
\newcommand{\seven}         {$\sqrt{s}~=~7$~Te\kern-.1emV\xspace}
\newcommand{\twoH}          {$\sqrt{s}~=~0.2$~Te\kern-.1emV\xspace}
\newcommand{\twosevensix}   {$\sqrt{s}~=~2.76$~Te\kern-.1emV\xspace}
\newcommand{\five}          {$\sqrt{s}~=~5.02$~Te\kern-.1emV\xspace}
\newcommand{\twosevensixnn} {$\sqrt{s_{\mathrm{NN}}}~=~2.76$~Te\kern-.1emV\xspace}
\newcommand{\fivenn}        {$\sqrt{s_{\mathrm{NN}}}~=~5.02$~Te\kern-.1emV\xspace}

\newcommand{\GeVc}          {Ge\kern-.1emV/$c$\xspace}
\newcommand{\MeVc}          {Me\kern-.1emV/$c$\xspace}
\newcommand{\TeV}           {Te\kern-.1emV\xspace}
\newcommand{\GeV}           {Ge\kern-.1emV\xspace}
\newcommand{\MeV}           {Me\kern-.1emV\xspace}
\newcommand{\GeVmass}       {Ge\kern-.1emV/$c^2$\xspace}
\newcommand{\MeVmass}       {Me\kern-.1emV/$c^2$\xspace}

\newcommand{\ZNA}    {\rm{ZNA}\xspace}
\newcommand{\ZNC}    {\rm{ZNC}\xspace}

\newcommand{\pip}    {\ensuremath{\pi^{+}}\xspace}
\newcommand{\pim}    {\ensuremath{\pi^{-}}\xspace}

\newcommand{\pbar}   {\ensuremath{\rm\overline{p}}\xspace}
\newcommand{\kzero}  {\ensuremath{{\rm K}^{0}_{\rm{S}}}\xspace}
\newcommand{\lmb}    {\ensuremath{\Lambda}\xspace}
\newcommand{\almb}   {\ensuremath{\overline{\Lambda}}\xspace}
\newcommand{\Om}     {\ensuremath{\Omega^-}\xspace}

\newcommand{\Pythia}{\textsc{Pythia}\xspace}
\newcommand{\Vzero}{\ensuremath{{\rm V}^{0}}}

\newcommand{\dd}{\ensuremath{{\rm d}}}

\newcommand{\pT}{\ensuremath{\pt}}
\newcommand{\pTv}{\ensuremath{p_{\rm T,\;\Vzero}}\xspace}
\newcommand{\pTj}{\ensuremath{p_{\rm T,\;jet}}\xspace}
\newcommand{\pTjch}{\ensuremath{\pTj^{\rm ch}}\xspace}

\newcommand{\hlab}{\ensuremath{\eta_{\rm lab}}\xspace}

\newcommand{\rhoch}{\ensuremath{\rho^{\rm ch}}\xspace}

\newcommand{\Rmat}{\ensuremath{R_{\rm match}}\xspace}
\newcommand{\Rcut}{\ensuremath{R_{\rm cut}}\xspace}
\newcommand{\Rvj}{\ensuremath{R(\Vzero,\;{\rm jet})}\xspace}

\newcommand{\kT}{\ensuremath{k_{\rm T}}\xspace}
\newcommand{\akT}{anti-$\kT$\xspace}

\newcommand{\Ajet}{\ensuremath{A_{\rm jet}}\xspace}

\newcommand{\rLK}{\ensuremath{(\lmb+\almb)/2\kzero}\xspace}

\newcommand{\rhoV}{\ensuremath{\rho^{\Vzero}}\xspace}
\newcommand{\drhoVdpT}{\ensuremath{\dd\rhoV/\dd\pT}\xspace}

\newcommand{\cent}   [2] {$#1$--$#2\%$}

\newcommand{\abs}[1]{\ensuremath{\left|#1\right|}}
\newcommand{\avg}[1]{\ensuremath{\left\langle#1\right\rangle}}

\begin{document}

\begin{titlepage}
\PHyear{2021}
\PHnumber{048}
\PHdate{22 March}

\title{Production of $\lmb$ and $\kzero$ in jets in \pPb collisions at \fivenn and \pp collisions at \seven}
\ShortTitle{Production of $\lmb$ and $\kzero$ in jets in \pPb collisions}

\Collaboration{ALICE Collaboration\thanks{See Appendix~\ref{app:collab} for the list of collaboration members}}
\ShortAuthor{ALICE Collaboration}

\begin{abstract}

The production of \lmb\ baryons and \kzero\ mesons ($\Vzero$ particles) was measured in \pPb collisions at \fivenn and \pp collisions at \seven with ALICE at the LHC.
The production of these strange particles is studied separately for particles associated with hard scatterings and the underlying event to shed light on the baryon-to-meson ratio enhancement observed at intermediate transverse momentum ($\pT$) in high multiplicity \pp and \pPb collisions.
Hard scatterings are selected on an event-by-event basis with jets reconstructed with the \akT\ algorithm using charged particles.
The production of strange particles associated with jets $\pTjch>10$ and $\pTjch>20$~\GeVc in \pPb collisions, and with jet $\pTjch>10$~\GeVc in pp collisions is reported as a function of $\pT$.
Its dependence on angular distance from the jet axis, $\Rvj$, for jets with $\pTjch > 10$~\GeVc in \pPb collisions is reported as well.
The $\pT$-differential production spectra of strange particles associated with jets are found to be harder compared to that in the underlying event and both differ from the inclusive measurements.
In events containing a jet, the density of the $\Vzero$ particles in the underlying event is found to be larger than the density in the minimum bias events.
The $\lmb/\kzero$ ratio associated with jets in \pPb collisions is consistent with the ratio in pp collisions and follows the expectation of jets fragmenting in vacuum. 
On the other hand, this ratio within jets is consistently lower than the one obtained in the underlying event and it does not show the characteristic enhancement of baryons at intermediate $\pT$ often referred to as ``baryon anomaly'' in the inclusive measurements.
\end{abstract}
\end{titlepage}

\setcounter{page}{2}

\section{Introduction}

High-energy heavy-ion collisions provide a unique opportunity to study properties of the hot and dense medium composed of deconfined partons, known as the quark--gluon plasma (QGP)~\cite{Shuryak:1984nq,Cleymans:1985wb,Bass:1998vz,Satz:2000bn,Jacak310,Muller:2012zq}.
A cross-over transition from hadronic matter to the QGP at zero baryochemical potential is expected to take place once the temperature reaches values of about $T_{\rm c} = 156$~\MeV based on quantum chromodynamics (QCD) calculations performed on a lattice~\cite{Borsanyi:2010cj,Bhattacharya:2014ara,Braun_Munzinger_2016}.
The measurements indicate that collisions of lead ions at the Large Hadron Collider (LHC) at a centre-of-mass energy per nucleon--nucleon collision of \twosevensixnn create conditions well above $T_{\rm c}$ at approximately zero baryochemical potential~\cite{Adam:2015lda}.

The interpretation of nucleus--nucleus (AA) collision results requires the understanding of results from smaller collision systems such as proton--proton (pp) or proton--nucleus (pA).
To separate initial state effects, linked to the use of nuclear beams or targets, from final-state effects, associated with the presence of hot and dense matter, particle production is compared in pp, pA, and AA reactions.
However, the measurements at the LHC in high-multiplicity pp and \pPb collisions have revealed unexpectedly strong long-range correlations of produced particles typical of \PbPb collisions~\cite{Khachatryan:2010gv,CMS:2012qk,Abelev:2012ola,Aad:2012gla,Aad:2013fja,Chatrchyan:2013nka,Acharya:2019icl,Adam:2015bka,Aad:2015gqa,Khachatryan:2015lva,Aaij:2015qcq}.
Measurements of identified light-flavour hadrons~\cite{ABELEV:2013wsa,Abelev:2013haa,Acharya:2018orn,Khachatryan:2014jra}, strange particles ~\cite{Chekanov:2006wz,ALICE:2017jyt,Khachatryan:2016yru,Sirunyan:2018toe}, and heavy-flavour particles~\cite{Aaij:2019lkm,Aaij:2019pqz} in small systems have also shown qualitatively similar features as in AA collisions~\cite{Adler:2003kg,Long:2004cf,Adams:2006wk,Fries:2008hs,Abelev:2013xaa,ABELEV:2013wsa}.
In particular, the baryon-to-meson yield ratio as a function of transverse momentum ($\pT$) shows a pronounced maximum at intermediate $\pT$ ($2$--$5$~\GeVc)~\cite{Chekanov:2006wz,Abelev:2013haa,Khachatryan:2016yru}.
The $\pT$ dependence of the ratio was discussed in terms of particle production within a common velocity field (collective flow)~\cite{Schnedermann:1993ws}, soft--hard parton recombination~\cite{Fries:2003vb} and high-energy parton shower (jet) hadronization at high $\pT$.
On the other hand, the jet suppression ascribed to the parton energy loss in the QGP observed in central AA collisions is not observed in \pPb collisions~\cite{Aad:2010bu,Chatrchyan:2012nia,Aad:2012vca,Abelev:2013kqa,Aad:2014bxa,Adam:2015hoa,Adam:2015xea,Khachatryan:2016xdg,Abelev:2014dsa,Aad:2016zif,Khachatryan:2015xaa,Adam:2016jfp}.
The measurements show that the impact of the initial-state nuclear effects such as shadowing and potential gluon saturation effects, e.g., Color Glass Condensate (CGC)~\cite{McLerran:2001sr,Salgado:2011wc}, or multiple scatterings and hadronic re-interactions in the initial and final states~\cite{Krzywicki:1979gv,Accardi:2007in} is small on the jet production in \pPb collisions.
To understand particle production mechanisms in small systems, the separation of particles produced in hard processes (jets) from those produced in the underlying event is important.
It allows one to investigate similarities and expose differences in particle production mechanisms in high-multiplicity \pp, \pPb events, and heavy-ion collisions.

In this letter, measurements of \kzero\ and \lmb\ (\almb), the $\Vzero$ particles, in \pPb collisions at \fivenn and \pp collisions at \seven are reported.
The production of $\Vzero$ particles is studied separately within the region associated with a hard parton scattering and the underlying event.
Hard scatterings are tagged by selecting a reconstructed jet with transverse momentum $\pTjch > 10$ or $20$~\GeVc using charged particles with the \akT algorithm~\cite{Cacciari:2008gp} and the resolution parameter $R=0.4$.
The baryon-to-meson ratio of $\Vzero$ particles associated with jets is reported as a function of particle transverse momentum and distance to the jet axis.
To contrast the strangeness production associated with a hard scattering and subsequent jet fragmentation with the production in the underlying event we report the ratio for the case of particles not associated with jets.
The $\pT$-differential ratio is also compared with a \Pythia~8 (version $8.2.43$; Tune $4$C)~\cite{Sj_strand_2015} simulation.

\section{Data analysis}

\subsection{The ALICE detector and data sample}

The ALICE apparatus consists of central barrel detectors covering the pseudorapidity interval $\abs{\hlab}<0.9$, a forward muon spectrometer covering $-4.0<\hlab<-2.5$, and a set of detectors at forward and backward rapidities used for triggering and event characterization.
Further information can be found in Ref.~\cite{Aamodt:2008zz}.
Tracking and particle identification in the context of this analysis are performed using the information provided by the Inner Tracking System (ITS)~\cite{Aamodt:2010aa} and the Time Projection Chamber (TPC)~\cite{Alme:2010ke}, which have full azimuthal coverage in the pseudorapidity interval $\abs{\hlab}<0.9$.
These central barrel detectors are located inside a large solenoidal magnet, which provides a magnetic field of $0.5$~T along the beam direction ($z$-axis in the ALICE reference frame).
The ITS is composed of six cylindrical layers of silicon detectors, with radial distances from the beam axis ranging from $3.9$~cm to $43.0$~cm.
The two innermost layers are equipped with Silicon Pixel Detectors (SPD) covering the pseudorapidity ranges of $\abs{\hlab} < 2.0$ and $\abs{\hlab} < 1.4$, respectively.
The two intermediate layers are made of Silicon Drift Detectors (SDD), while Silicon Strip Detectors (SSD) equip the two outermost layers.
The high spatial resolution of the silicon sensors, together with the low material budget (on average $7.7\%$ of a radiation length for tracks crossing the ITS perpendicularly to the detector surfaces, i.e., $\hlab=0$) and the small distance of the innermost layer from the beam pipe, allow for the measurement of the track impact parameter $d_{\rm DCA}$ in the transverse plane.
The $d_{\rm DCA}$ is defined by the distance of closest approach (DCA) of the track to the primary vertex in the plane transverse to the beam direction, and is measured with a resolution better than $75$~$\mu$m for transverse momenta $\pT>1$~\GeVc, including the contribution from the primary vertex position resolution~\cite{Aamodt:2010aa}.
At larger radii ($85<r<247$~cm), the $500$~cm long cylindrical TPC provides track reconstruction with up to $159$ three-dimensional space points per track, as well as particle identification via the measurement of the specific energy deposit $\dEdx$ in the gas.
The overall $\pT$ resolution given by combining ITS and TPC information is typically $1\%$ for momenta of $1$~\GeVc and $7\%$ for momenta of $10$~\GeVc~\cite{Aamodt:2010my}.

The data sample used in this analysis was recorded by the ALICE detector~\cite{Aamodt:2008zz} during the LHC \pPb run at \fivenn and \pp\ run at \seven in $2013$ and $2010$, respectively.
Because of the $2$-in-$1$ magnet design of the LHC~\cite{Evans:2008zzb}, the energies of the two beams are not independent and their ratio is fixed to be equal to the ratio of the charge-to-mass ratios of each beam.
Consequently, for \pPb collisions, the nucleon--nucleon centre-of-mass system is shifted in rapidity by $\Delta y_{\rm NN}=0.465$ in the direction of the proton beam.
In the analyzed data sample the Pb beam circulated in the ``counter-clockwise'' direction travelling from negative to positive rapidity in the laboratory reference frame.
The setup of the detector, trigger, and the analysis strategy is identical in both collision systems unless explicitly stated otherwise.

The data samples presented in this letter were recorded using the minimum bias trigger implemented by the VZERO detector~\cite{Abbas:2013taa}.
The VZERO system consists of two arrays of $32$ scintillator tiles each, placed around the beam vacuum pipe on either side of the interaction region covering the pseudorapidity intervals $2.8 < \hlab < 5.1$ (VZERO-A) and $-3.7 < \hlab < -1.7$ (VZERO-C).
In addition, in \pPb collisions, two neutron Zero Degree Calorimeters (ZDCs), located at $+112.5$~m (\ZNA) and $-112.5$~m (\ZNC) from the interaction point, are used in the offline event selection for rejecting of beam-background events, exploiting the correlation between the arrival times measured in \ZNA and \ZNC.
In pp collisions, a logical OR between the requirement of at least one hit in the SPD and a hit in one of the two VZERO scintillator arrays is used for event selection.
In \pPb collisions, a coincidence of signals in both VZERO-A and VZERO-C is required to remove contamination from single-diffractive and electromagnetic events~\cite{ALICE:2012xs}.
The events are further selected to require a reconstructed vertex within $10$~cm ($\abs{v_{z}}<10$~cm) of the nominal centre of the detector along the beam axis and vertices built from the SPD tracklets, which are the short track segments measured with SPD, and from the tracks measured with combined information from ITS and TPC are compatible.
The fraction of events with the vertex selection criteria is about $98.2\%$ of all triggered events.
In total, about $96\times 10^{6}$ ($177\times 10^{6}$) events, corresponding to an integrated luminosity of $\mathcal{L}\approx 46$~$\mu$b$^{-1}$ ($2.9$~nb$^{-1}$), are used in the analysis of the \pPb (pp) data sample.

\subsection{Charged-particle and jet reconstruction}

The charged-particle reconstruction and jet reconstruction in this letter follow the approach described in detail in Refs.~\cite{Adam:2015hoa,ALICE:2014dla}.
Here only a brief review of the most relevant points is given.
Charged-particle tracks, reconstructed in the ITS and the TPC with $\pT > 0.15$~\GeVc and within the TPC acceptance $\abs{\hlab}<0.9$ that satisfy a DCA requirement $d_{\rm DCA} < 2.4$~cm, are used as input to the jet reconstruction.
The azimuthal distribution of these tracks is not completely uniform due to inefficient regions in the SPD.
This is compensated by considering in addition tracks with less than three reconstructed track points in the ITS or no points in the SPD.
To improve the momentum resolution for those tracks, the primary vertex is used as an additional constraint in the track fitting.
This approach yields a uniform tracking efficiency within the acceptance.
These complementary tracks constitute approximately $4.3\%$ and $5\%$ of the overall used track sample in \pPb and pp collisions, respectively.
The efficiency for charged-particle detection, including the effect of tracking efficiency as well as the geometrical acceptance, is $70\%$ ($60\%$) at $\pT = 0.15$~\GeVc and increases to $85\%$ ($87\%$) at $\pT = 1$~\GeVc and above for \pPb (pp) collisions.

The jets are reconstructed using the \akT\ algorithm~\cite{Cacciari:2008gp} from the FastJet package~\cite{Cacciari:2011ma,Cacciari:2005hq} with resolution parameter $R=0.4$.
Only those jets for which the jet-axis is found within the acceptance window $\abs{\hlab}<0.35$ are used in this analysis.
This condition ensures the jet cone is fully overlapping with the acceptances of both charged-particle tracks ($\abs{\hlab}<0.9$) and the \Vzero\ particles ($\abs{\hlab}<0.75$, as explained in detail in section~\ref{sec:V0rec}).
The jet transverse momentum is calculated with FastJet using the $\pt$ recombination scheme.

In general, the transverse-momentum density of the background ($\rhoch$), originating from the underlying event and/or pile-up, contributes to the jet energy reconstructed by the jet finder.
The correction of the jet-energy scale accounting for the background contribution can be estimated on an event-by-event basis using the median of the transverse momentum density of all the clusters reconstructed with the \kT\ algorithm~\cite{Cacciari:2008gn}.
In pp and \pPb collisions, an estimate adequate for the more sparse environment than \PbPb collisions is employed by scaling $\rhoch$ with an additional factor to account for event regions without particles~\cite{Adam:2015hoa}. The resulting mean of the background $\pT$ density in \pPb collisions is $\avg{\rhoch}=1.02$~\GeVc~rad$^{-1}$ (with negligible statistical uncertainty) for unbiased events and $\avg{\rhoch}=2.2~\pm~0.01$~\GeVc~rad$^{-1}$ for events containing a jet with uncorrected transverse momentum $\pTj^{\rm ch,\;raw} >20$~\GeVc~\cite{Adam:2015hoa}.
In pp collisions, the background density is around $1$~\GeVc~rad$^{-1}$ and not subtracted on a jet-by-jet basis but the related uncertainty on the jet $\pt$ scale is absorbed into the systematic uncertainty.

The jet finding efficiency, which encodes the effects of single-particle momentum resolution and reconstruction efficiency on the jet reconstruction, is estimated using a \Pythia~6~\cite{Sjostrand:2006za}~$+$~GEANT~3~\cite{Brun:1994aa} simulation by comparing the generated jets to reconstructed ones and found to be larger than $96\%$ in the considered momentum range ($\pTjch > 10$~\GeVc).

\subsection{Reconstruction of \Vzero\ particles}%
\label{sec:V0rec}

The \Vzero\ particles, \kzero\ and \lmb\ (\almb), are identified by taking advantage of the characteristics of their weak decay topologies in the channels $\kzero\to\pip\pim$ and $\lmb\ (\almb)\to{\rm p}\pim\ (\pbar\pip)$, which have branching ratios of $69.2\%$ and $63.9\%$, respectively~\cite{Tanabashi:2018oca}.
The reconstruction and the selection criteria of the \Vzero\ particles follow the analysis in Ref.~\cite{Abelev:2013haa} with the exception of the rapidity selection of the particles and their decay products.
The decay products of the \Vzero\ particles, $\pi^{\pm}$ and p ($\overline{\rm p}$), are identified in the central barrel with the TPC using the specific energy loss $\dEdx$ in the gas by measuring up to $159$ samples per track with a resolution of about $6\%$~\cite{Abelev:2014ffa}.
Since the \Vzero\  daughter tracks are displaced from the primary vertex and tracks in the jet are selected by criteria optimised for particles produced at the primary vertex, only about $0.1\%$ of the \Vzero\ daughter tracks contribute to the charged-particle jet reconstruction.
The \Vzero\ decay daughter tracks are selected in the acceptance window $\abs{\hlab}<0.8$ following the criteria used in the inclusive analysis~\cite{Aamodt:2011zza,Abelev:2013haa}.
To avoid the fiducial effect, only the \Vzero\ candidates found in $\abs{\hlab}<0.75$ are retained.
This ensured that the reconstruction efficiency is approximately constant throughout the selected pseudorapidity range.
The topological selection of \Vzero\ candidates within the kinematic range of this analysis yields almost background-free invariant mass spectra with the lowest signal-to-background ratio among all of the \Vzero\ particles still exceeding $10$.
The $\pT$-differential yields of the \Vzero\ particles are extracted using the invariant-mass method, described in Ref.~\cite{Abelev:2013haa}, where the combinatorial background is interpolated from the side bands defined in terms of the mass peak width $\sigma$ in intervals $[-12\sigma,-6\sigma]$ and $[6\sigma,12\sigma]$ with respect to the mean of the peak.

\subsection{Matching of \Vzero\ particles to jets and underlying event}
\label{sec:c05V0JetMat}

To obtain the yield of \Vzero\ particles within a jet cone, the \Vzero\ particles are selected based on their distance from the jet centroid in the pseudorapidity ($\hlab$) and azimuthal angle ($\varphi$) plane
\begin{equation}
\Rvj=\sqrt{\left(\hlab^{\rm jet} - \hlab^{\Vzero}\right)^{2} + \left(\varphi^{\rm jet} - \varphi^{\Vzero}\right)^{2}}.
\end{equation}
A~$\Vzero$ particle with a radial distance from a given jet $\Rvj<\Rmat$ is considered matched to the jet and referred to as the ``$\Vzero$ inside the jet cone" (JC $\Vzero$).
In \pPb collisions the probability for a particle with $\pT>0.5$~\GeVc to lie in the overlapping region of two different jets with $\pTjch > 10$~\GeVc is less than $1\%$ and in these cases the higher-energy jet is preferred.
Moreover, removal of the events with the same particle matching to two or more jets did not alter the result of the analysis.
The procedure for extracting the yield of \Vzero\ particles, associated with a jet within a cone defined by $\Rmat$, can be summarised as follows.
For each $\pT$ interval the JC \Vzero\ yield is extracted using the invariant mass technique, where the combinatorial background is interpolated from the side bands.
Then the raw JC \Vzero\ yield is corrected for the contribution of particles from the underlying event (the UE $\Vzero$).

Conceptually, the UE \Vzero{} particles represent the particles that are not associated with the hard scatterings tagged by the charged jets considered in this analysis.
To extract the UE \Vzero\ yield several estimators were investigated: i) an \emph{outside cone} (OC) selection, composed of the \Vzero\ particles that satisfy the condition of $\Rvj > \Rcut$ (e.g. $\Rcut=0.6$) within events containing a jet; ii) the \emph{perpendicular cone} (PC) selection, composed of the \Vzero\ particles found in a range with radius $R=0.4$ in $\eta$ and $\varphi$ space perpendicular to the jet axis  at the same $\eta$; and iii) the \emph{non-jet event} (NJ) selection, composed of the \Vzero\ particles found in events that do not contain a jet with $\pTjch>5$~\GeVc.

In practice, a useful quantity for performing the subtraction of the non-jet contribution of the \Vzero\ particles is their density per unit area
\begin{equation}
\rhoV(\pT) = N^{\Vzero}(\pT) / A^{\Vzero},
\label{eq:defv0rho}
\end{equation}
where $N^{\Vzero}$ is the number of $\Vzero$ particles and $A^{\Vzero}$ is the acceptance in pseudorapidity and azimuthal angle.
Consequently, the number of the UE \Vzero\ particles within a jet cone can be calculated as $N=\rhoV\Ajet$ for each estimator separately.
The jet area $\Ajet = \pi \Rmat^{2}$ is considered in this analysis.
In general the density of \Vzero\ particles within jets can be defined as
\begin{equation}
\rhoV_{\rm JC} - \rhoV_{\rm UE},
\label{eq:defJEV0s}
\end{equation}
where UE can be any of the OC, PC, or NJ background estimators.
In this analysis, PC is chosen as the default background estimator, while OC and NJ are used to quantify the systematic uncertainty.

\subsection{Corrections for finite \Vzero\ reconstruction efficiency and feed-down}
\label{sec:c05V0EffiMC}

The reconstruction efficiencies of \Vzero\ particles are estimated using the DPMJET~\cite{Roesler:2000he} and \Pythia~$6$~\cite{Sjostrand:2006za} Monte Carlo generators in \pPb and \pp collisions, respectively, with the same selection criteria as in the data except the daughter track particle identification with $\dEdx$ in the TPC (see more details in~\cite{Abelev:2013haa}).
These simulations are based on the GEANT~$3$ transport code~\cite{Brun:1994aa} for the detector description and response.

\begin{figure}[ht]
\begin{center}
\includegraphics[width=.9\textwidth]{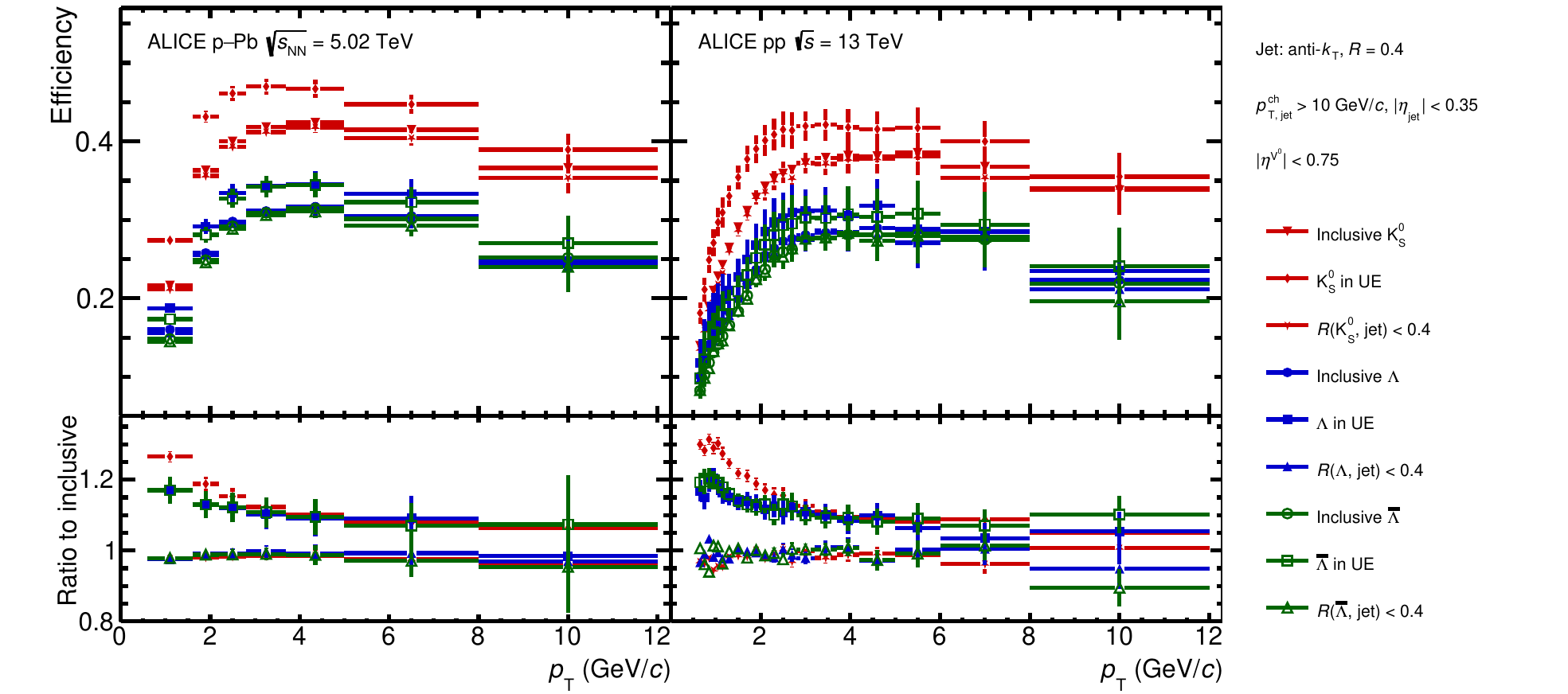}
\caption{Reconstruction efficiency of \Vzero\ particles in \pPb\ collisions at \fivenn (left panel) and in pp collisions at \seven (right panel) for three selection criteria: inclusive, within $\Rvj<0.4$ and \Vzero{}s in UE (upper panels) and the ratio relative to inclusive selection (lower panels).
UE \Vzero{}s are estimated with the OC estimator ($\Rvj>0.6$) in \pPb collisions and with the PC estimator in pp collisions.}
\label{fig:c02EffiIncV0s}
\end{center}
\end{figure}

Due to differences in the experimental acceptance for \Vzero\ particles associated with jets and those extracted through the various estimators of the underlying event, the efficiencies of \Vzero\ particles are estimated separately for every case.
Figure~\ref{fig:c02EffiIncV0s} shows the reconstruction efficiencies for inclusive \Vzero{}s and those for JC \Vzero{}s with $\Rvj<0.4$ and UE \Vzero{}s.
The UE \Vzero{}s are estimated with the OC estimator with $\Rvj>0.6$ in \pPb collisions and with the PC estimator in pp collisions.
In particular, for $\Rvj<0.4$ the efficiency at $\pT<2$~\GeVc is about $20\%$ larger than in the inclusive case while it approaches the inclusive case at higher $\pT$.
This is due to the fact that the $\eta$-differential reconstruction efficiency of \Vzero\ particles decreases with $\abs{\hlab}$ and the pseudorapidity distribution of \Vzero\ particles matched with jets is narrower than that of inclusive ones.
This results in a higher $\eta$-integrated efficiency of JC \Vzero{}s than inclusive \Vzero{}s.
This effect is more pronounced at low $\pT$.

The $\pT$-differential yields of \lmb\ and \almb\ reconstructed for JC and UE selections are also corrected for the feed-down from the decays of $\Xi^{0}$ and $\Xi^{-}$ particles and their respective anti-particles.
The $\Xi$ production in jets is estimated based on measurements of the multi-strange baryons and their decays at high $\pT$ performed in pp collisions~\cite{Abelev:2012jp} and extrapolated to low $\pT$ using the \Pythia~$8$ event generator.
The applied correction amounts to $15\%$ and is independent of the \lmb\ and \almb\ momenta.
Conversely, the \lmb\ yields are not corrected for the feed-down from $\Om$ baryons as this contribution is negligible compared to the systematic uncertainties of the present measurement.
Since \lmb\ from non-weak decays of the $\Sigma^{0}$ and $\Sigma^{*}(1385)$ family cannot be distinguished from the direct ones, the identified \lmb\ yield includes these contributions~\cite{Aamodt:2011zza}.

\subsection{Systematic uncertainties}
\label{sec:uncertainties}

The main sources of systematic uncertainty in the \Vzero\ particle reconstruction are uncertainties on the material budget ($4\%$), the track selection (up to $4\%$), feed-down correction for the \lmb\ ($5\%$ for $\pT<3.7$~\GeVc and $7\%$ for $\pT>3.7$~\GeVc), proper lifetime selection criteria (up to $\sim 4\%$), and topological selections depending on transverse momentum and particle species (up to $1.6\%$).
The systematic uncertainties on the extracted yields for \kzero\ mesons and \lmb\ and \almb\ baryons in \pPb and pp collisions are reported as point-to-point uncertainties in Table~\ref{tab:systpPb} and Table~\ref{tab:systpp}.
The ``negl.'' in the table denotes an uncertainty of less than $0.1\%$.
The total uncertainty on the yields is calculated by adding the individual uncertainties on track selection, material budget, feed-down corrections and the listed \Vzero\ selections in quadrature.

\begin{table}[ht]
\centering
\caption{Relative systematic uncertainties in percent for $\kzero$, $\lmb$, and $\almb$ in p--Pb collisions at \fivenn.
The right three columns in the last two rows represent the uncertainties of $\lmb + \almb$.
For each particle, the reported values correspond to the uncertainties at $\pT=0.6$, $2$, and $10$~\GeVc.
See text for details.}
\begin{tabularx}{\textwidth}{@{} lCCCCCCCCC @{}}
\toprule
~ & \multicolumn{3}{c}{$\kzero$} & \multicolumn{3}{c}{$\lmb$} & \multicolumn{3}{c}{$\almb$} \\
\cmidrule(lr{.5em}){2- 4}
\cmidrule(lr{.5em}){5- 7}
\cmidrule(lr{.5em}){8-10}
Particle identification      & \multicolumn{3}{c}{$<1$}  & negl. & negl. & $3.2$ & negl. & negl. & $2.7$ \\
Track selection              & negl. & negl. & $1.5$     & negl. & $1$   & $1.7$ & negl. &  $1$  & $2$   \\
Topological selection        & $0.3$ & $1.1$ & $0.7$     & $0.7$ & $1.6$ & $0.7$ & $0.6$ & $1.6$ & $0.5$ \\
Proper lifetime              & \multicolumn{3}{c}{negl.} & $2.9$ & $3.7$ & negl. & $2.6$ & $3.5$ & negl. \\
Competing $\Vzero$ selection & \multicolumn{3}{c}{$<1$}  & negl. & negl. & $3.4$ & negl. & negl. & $1.3$ \\
Signal extraction            & $1.8$ & $4.2$ & $2.4$     & $2.1$ & $1.8$ & $1.3$ & $1.3$ & $1.8$ & $1.3$ \\
\cmidrule(lr{.5em}){5-10}
Jet $\pt$ scale & $1.4$ & $5.4$ & $37.1$ & \multicolumn{2}{r}{$3.2$} & \multicolumn{2}{c}{$4.2$} & \multicolumn{2}{l}{$41.1$} \\
UE subtraction  & $21$  & $9.6$ & $1.2$  & \multicolumn{2}{r}{$40$}  & \multicolumn{2}{c}{$24$}  & \multicolumn{2}{l}{$0.6$}  \\
\bottomrule
\end{tabularx}
\label{tab:systpPb}
\end{table}

\begin{table}[ht]
\centering
\caption{Relative systematic uncertainties in percent for $\kzero$, $\lmb$, and $\almb$ in pp collisions at \seven.
The right three columns in the last two rows represent the uncertainties of $\lmb + \almb$.
For each particle, the reported values correspond to the uncertainties at $\pT=0.6$, $2$ and, $10$~\GeVc.
See text for details.}
\begin{tabularx}{\textwidth}{@{} lCCCCCCCCC @{}}
\toprule
~ & \multicolumn{3}{c}{$\kzero$} & \multicolumn{3}{c}{$\lmb$} & \multicolumn{3}{c}{$\almb$} \\
\cmidrule(lr{.5em}){2- 4}
\cmidrule(lr{.5em}){5- 7}
\cmidrule(lr{.5em}){8-10}
Particle identification      & negl. & negl. & $2.3$     & negl. & negl. & $5$      & negl. & negl. & $2.5$  \\
Track selection              & $1.8$ & $3.6$ & $2.8$     & $1.7$ & $4$   & $3.6$    & $1.5$ & $4.3$ & $2.5$  \\
Topological selection        & $2.4$ & negl. & negl.     & $3.5$ & $0.7$ & $5.1$    & $3.7$ & $1.5$ & $2.5$  \\
Proper lifetime              & \multicolumn{3}{c}{negl.} & $2.6$ & $3$   & negl.    & $2.7$ & $2.5$ & negl.  \\
Competing $\Vzero$ selection & \multicolumn{3}{c}{$<1$}  & negl. & negl. & $9.5$    & negl. & negl. & $10.6$ \\
Signal extraction            & negl. & negl. & $1.4$     & \multicolumn{3}{c}{$<1$} & negl. & negl. & $1.3$  \\
\cmidrule(lr{.5em}){5-10}
Jet $\pt$ scale & $0.5$ & $1$   & $9.9$ & \multicolumn{2}{r}{$5.3$} & \multicolumn{2}{c}{$2$}    & \multicolumn{2}{l}{$10$}  \\
UE subtraction  & $4.8$ & $3.6$ & negl. & \multicolumn{2}{r}{$8.9$} & \multicolumn{2}{c}{$11.2$} & \multicolumn{2}{l}{negl.} \\
\bottomrule
\end{tabularx}
\label{tab:systpp}
\end{table}

\textbf{Particle identification (PID).} The uncertainty due to the particle identification is estimated by varying the selection criteria of the $\dEdx$ in the TPC from a default $5\sigma$ to $4$, $6$ and $7$ standard deviations from the nominal $\dEdx$ for pions and protons normalised to the detector resolution.

\textbf{Track selection.} The uncertainty originating from the track selection is estimated by repeating the analysis with an increased number of required TPC space points per track by about $7\%$ and $15\%$ from the nominal requirement of $70$ points.

\textbf{Topological selection.} The uncertainty associated with the topological selection of the \Vzero\ candidates (the two-dimensional decay radius, daughter track DCA to primary vertex, DCA of \Vzero\ daughters, and cosine of the pointing angle) is obtained by varying the parameters of the selections for each of the \Vzero\ species separately as described in detail in Ref.~\cite{Abelev:2013haa}.

\textbf{Proper lifetime selection.} The uncertainty due to the selection on the proper lifetime of \Vzero\ candidates, defined as the product of the mass $m_{0}$, decay length $L$, and the inverse of the particle momentum $p$ ($m_{0}Lc/p<20$~cm for \kzero\ and $m_{0}Lc/p<30$~cm for \lmb\ and \almb), is obtained by redoing the analysis with different selection criteria ($12$ and $40$~cm for \kzero\ and $20$ and $40$~cm for \lmb\ and \almb).

\textbf{Competing \Vzero\ selection.} The invariant mass of each candidate can be calculated either under the $\kzero$ or the $\lmb$ ($\almb$) mass hypothesis.
A $\kzero$ candidate is rejected if its invariant mass under the hypothesis of a $\lmb$ or $\almb$ lies in the window of $\pm 10$~\MeVmass around the mass of the $\lmb$ or $\almb$, and a $\lmb$ ($\almb$) candidate is rejected if its invariant mass under the $\kzero$ hypothesis lies in the window of $\pm 5$~\MeVmass around the $\kzero$ mass.
To assess the uncertainty related to this selection the analysis is repeated varying the invariant mass window of $3$ and $6$~\MeVmass for \kzero\ and with no rejection for \lmb\ ($\almb$) baryons.

\textbf{Underlying event subtraction.} Two main sources of uncertainties originating from the mis-association of \Vzero\ particles with the UE are considered: i) the \Vzero\ particle is found outside the selected jet and is classified as an UE particle; however, it may have originated from a physical jet outside the fiducial acceptance of jets considered in the analysis and/or from a \emph{true} low-$\pT$ jet, below the considered thresholds; and ii) the \Vzero\ particle originates from a true high-$\pT$ jet; however, due to the finite detector efficiency the jet has not been reconstructed above the considered $\pT$ threshold.

The uncertainty on the UE \Vzero\ density is estimated using the OC and NJ selections as alternatives for the density calculation, since the former is sensitive to particles outside the jet cone but originating from a physical jet and the latter is sensitive to those signals contributing to the UE due to the finite detector efficiency.
The standard deviation of the difference of the reconstructed \Vzero\ yields in OC and NJ is included as an additional systematic uncertainty on the density of particles within the jets.
In \pPb collisions the uncertainty is largest for low-momentum particles ($\pT < 2$~\GeVc) reaching up to $20\%$ ($40\%$) for $\kzero$ ($\lmb$) but drops rapidly with $\pT$ to negligible values for $\pT > 6$~\GeVc.
For pp collisions the trend of the uncertainty is similar to the trend seen in \pPb, however the magnitude is smaller, reaching values up to $5\%$ ($9\%$) for $\kzero$ ($\lmb$).

\textbf{Jets $\pt$ scale.} The systematic uncertainty originating from the selection of the jet $\pT$ is estimated by varying the jet $\pT$ around the chosen thresholds of $10$ and $20$~\GeVc by $2$~\GeVc.
This variation accounts for jet resolution effects due to detector effects and the fluctuations of the event background density as reported in Ref.~\cite{Adam:2015hoa}.
For jets with $\pTjch>10$~\GeVc at low momenta ($\pTv <2$~\GeVc) it reaches up to $10\%$, while it is about $20\%$ for jets of $\pTjch>20$~\GeVc.
It remains almost constant at about $3\%$ for $\pTv > 2$~\GeVc for jets $\pTjch>10$~\GeVc and about $5\%$ for jets $\pTjch>20$~\GeVc.

\textbf{Uncertainty of the $\rLK$ ratio.} The uncertainties on \Vzero\ yields, material budget and feed-down correction are propagated to the ratio quadratically.
The uncertainties related to the jet $\pT$ and UE estimation are obtained by calculating the deviation of ratios between the default analysis and various selection criteria.
Table~\ref{tab:systR} shows the point-to-point relative systematic uncertainties on the $\rLK$ ratio reconstructed within $R=0.4$ jets with $\pTjch > 10$~\GeVc (left column) and $\pTjch > 20$~\GeVc (middle column) in \pPb collisions.
For $\pTjch > 20$~\GeVc, the total uncertainty is about $16\%$ and is largely independent of particle $\pT$ with the largest contribution of \cent{8}{9} originating from the uncertainty on the \Vzero\ reconstruction.
The relative systematic uncertainties on the $\rLK$ ratio for pp collisions are shown in the right column in table~\ref{tab:systR}.

\begin{table}[!t]
\centering
\caption{Relative systematic uncertainties in percent for the $(\lmb+\almb)/2\kzero$ ratio of the spectrum of $\kzero$ and $\lmb$ ($\almb$) for $\pTjch>10$ and $20$~\GeVc in p--Pb collisions at \fivenn, and for $\pTjch>10$~\GeVc in pp collisions at \seven.
For each case, the reported values correspond to the uncertainties at $\pT=0.6$, $2$, and $10$~\GeVc.}
\begin{tabularx}{\textwidth}{@{} lCCCCCCCCC @{}}
\toprule
~ & \multicolumn{6}{c}{p--Pb} & \multicolumn{3}{c}{pp} \\
\cmidrule(lr{.5em}){2-7}
& \multicolumn{3}{c}{$\pT>10$~\GeVc} & \multicolumn{3}{c}{$\pT>20$~\GeVc} & \multicolumn{3}{c}{$\pT>10$~\GeVc} \\
\cmidrule(lr{.5em}){2 -4}
\cmidrule(lr{.5em}){5 -7}
\cmidrule(lr{.5em}){8-10}
$\Vzero$ reconstruction & $8.3$  & $8.9$  & $9.2$ & $8.8$  & $8.1$ & $9.2$ & $4.3$ & $4.8$ & $11.9$ \\
Jet $\pt$ scale         & $1.5$  & $4.2$  & $3.3$ & $9.2$  & $2.4$ & $7.4$ & $1.6$ & $2$   & $0.7$  \\
UE subtraction          & $26.3$ & $10.3$ & negl. & $21.1$ & $7.1$ & negl. & $7.5$ & $8.2$ & negl.  \\
\bottomrule
\end{tabularx}
\label{tab:systR}
\end{table}

\section{Results}%
\label{sec:Results}

In the following, results for $\Vzero$ particles with four different selections are discussed.
Their labels in the figures are defined as follows:
i) $\Vzero$s obtained from the unbiased events without any jet veto are labelled as ``Inclusive'' particles;
ii) $\Vzero$s matched to jets in a cone with a radius of $0.4$ are labelled as particles within ``$\Rvj < 0.4$'', the remaining underlying event background is not subtracted from this sample;
iii) the label ``$\Vzero$s in jets'' refers to $\Vzero$s produced in jets obtained by subtracting the underlying event background from the previous sample;
iv) $\Vzero$s from the underlying event estimated in cones perpendicular to the jet axis are labelled as "Perp. cone" particles.

\begin{figure}[ht]
\centering
\includegraphics[width=.74\textwidth]{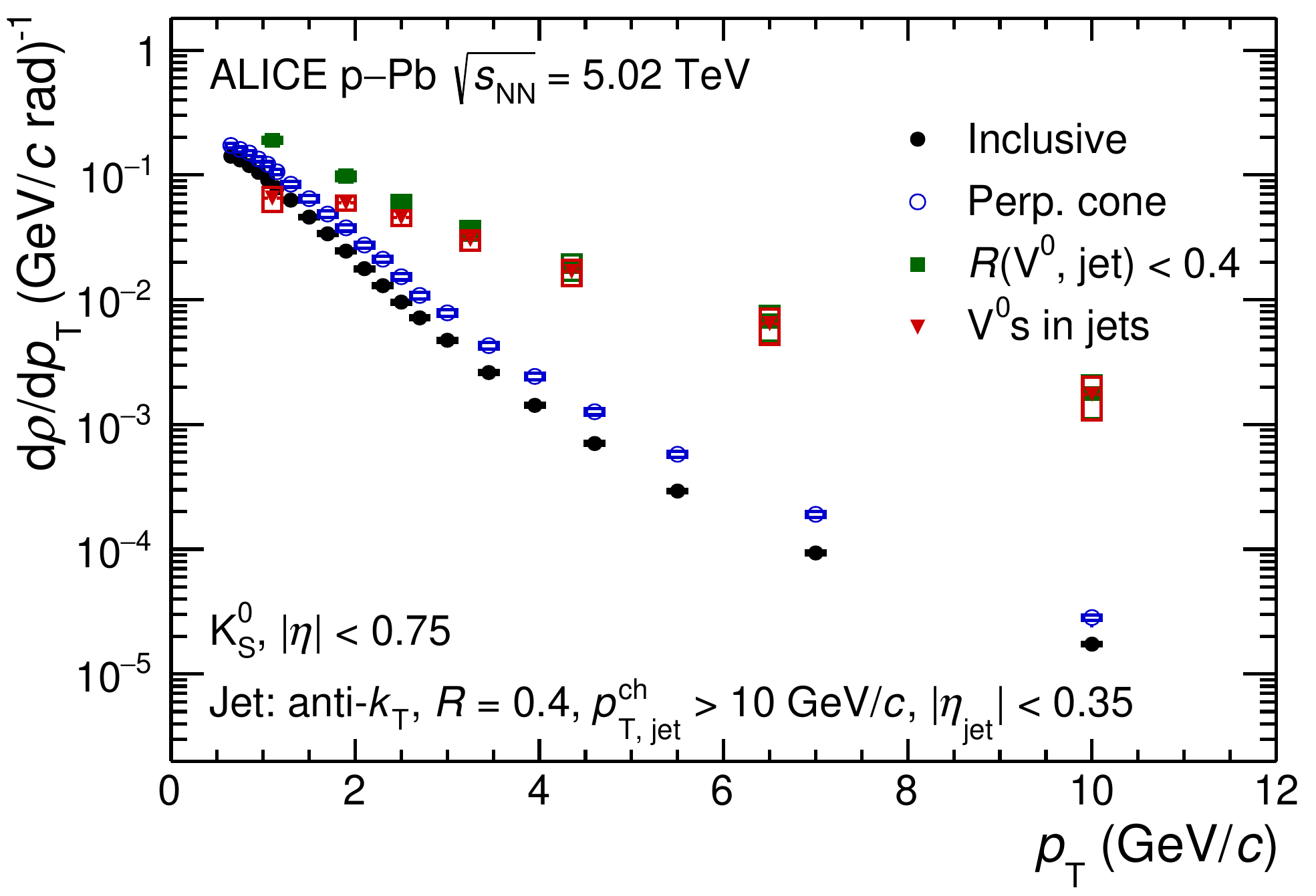}  \\
\includegraphics[width=.74\textwidth]{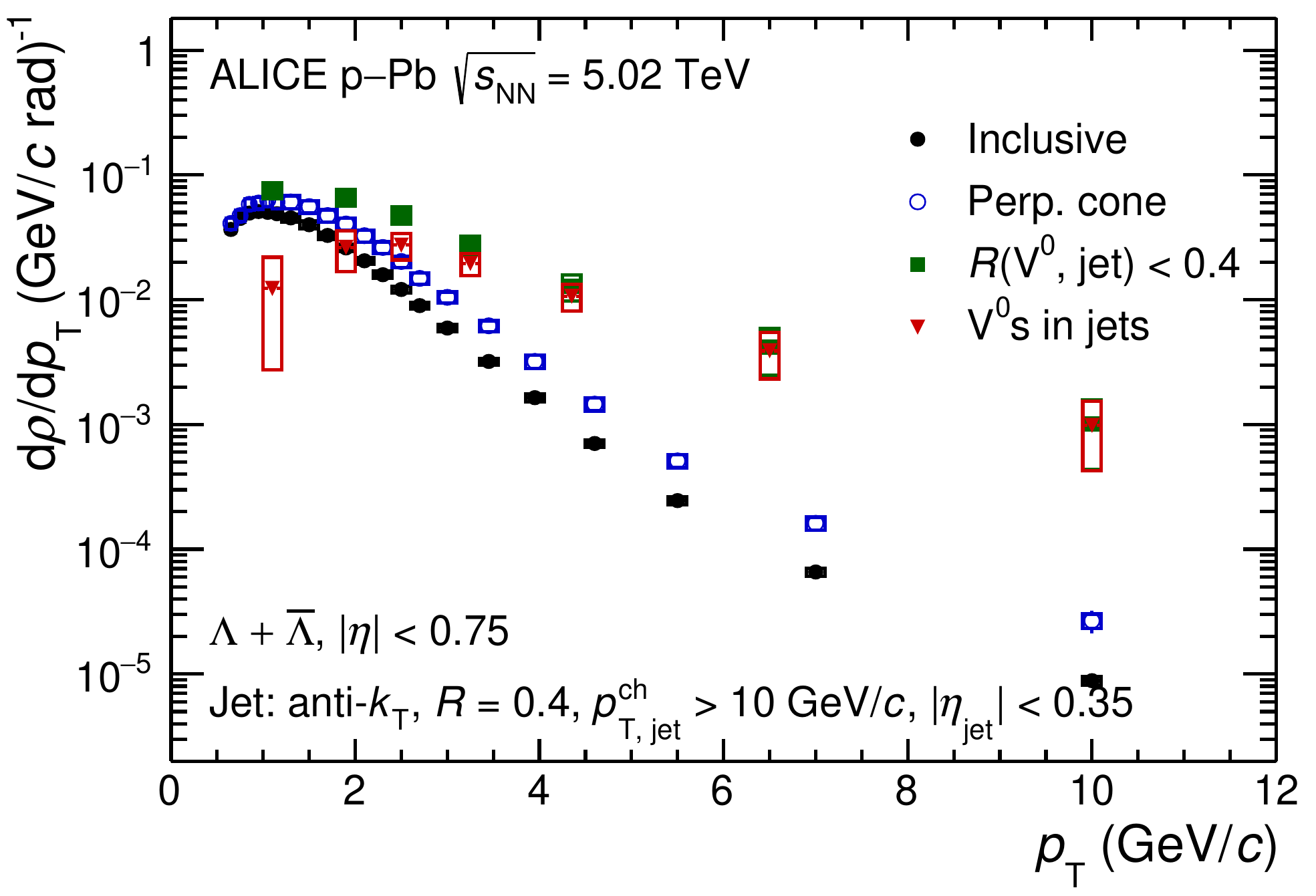}
\caption{The $\pT$-differential density of particles $\drhoVdpT$ (see Eq.~\eqref{eq:defv0rho}) in \pPb collisions at \fivenn for \kzero\ (upper panel) and the sum of \lmb\ and \almb\ (lower panel).
The density is shown for three selection criteria: inclusive particles from minimum bias events (black full circle), particles associated with the underlying event production estimated with PC selection (blue open circle, labelled as ``Perp. cone" in the figure), JC $\Vzero$s with $\Rvj < 0.4$ (green full square).
The density distribution of $\Vzero$s in jets with UE background subtracted (defined by Eq.~(\ref{eq:defJEV0s})) is shown as the red full triangle.
Statistical uncertainties and systematic uncertainties are shown as vertical bars and open boxes, respectively.}
\label{fig:rhov0pPb}
\end{figure}

\begin{figure}[ht]
\centering
\includegraphics[width=.74\textwidth]{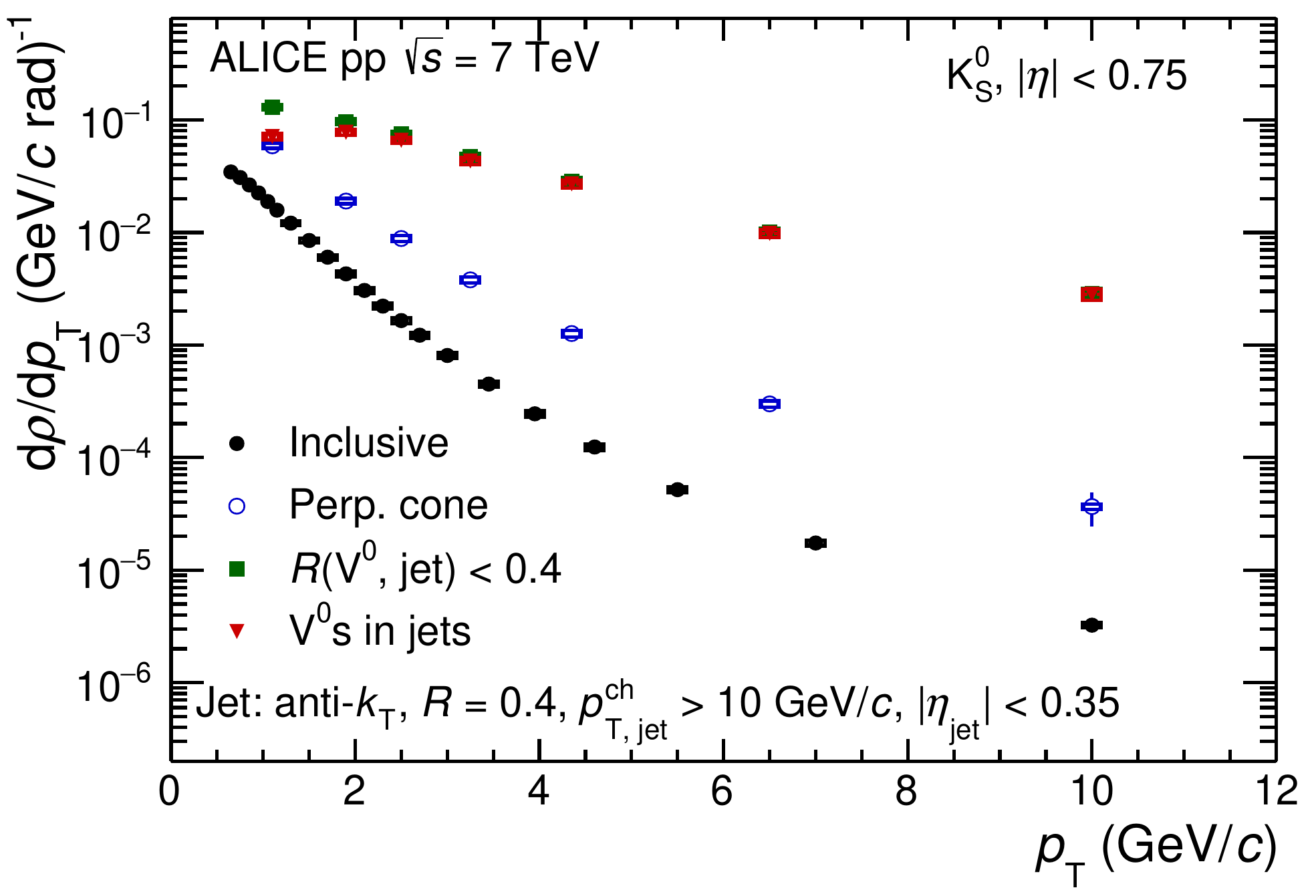} \\
\includegraphics[width=.74\textwidth]{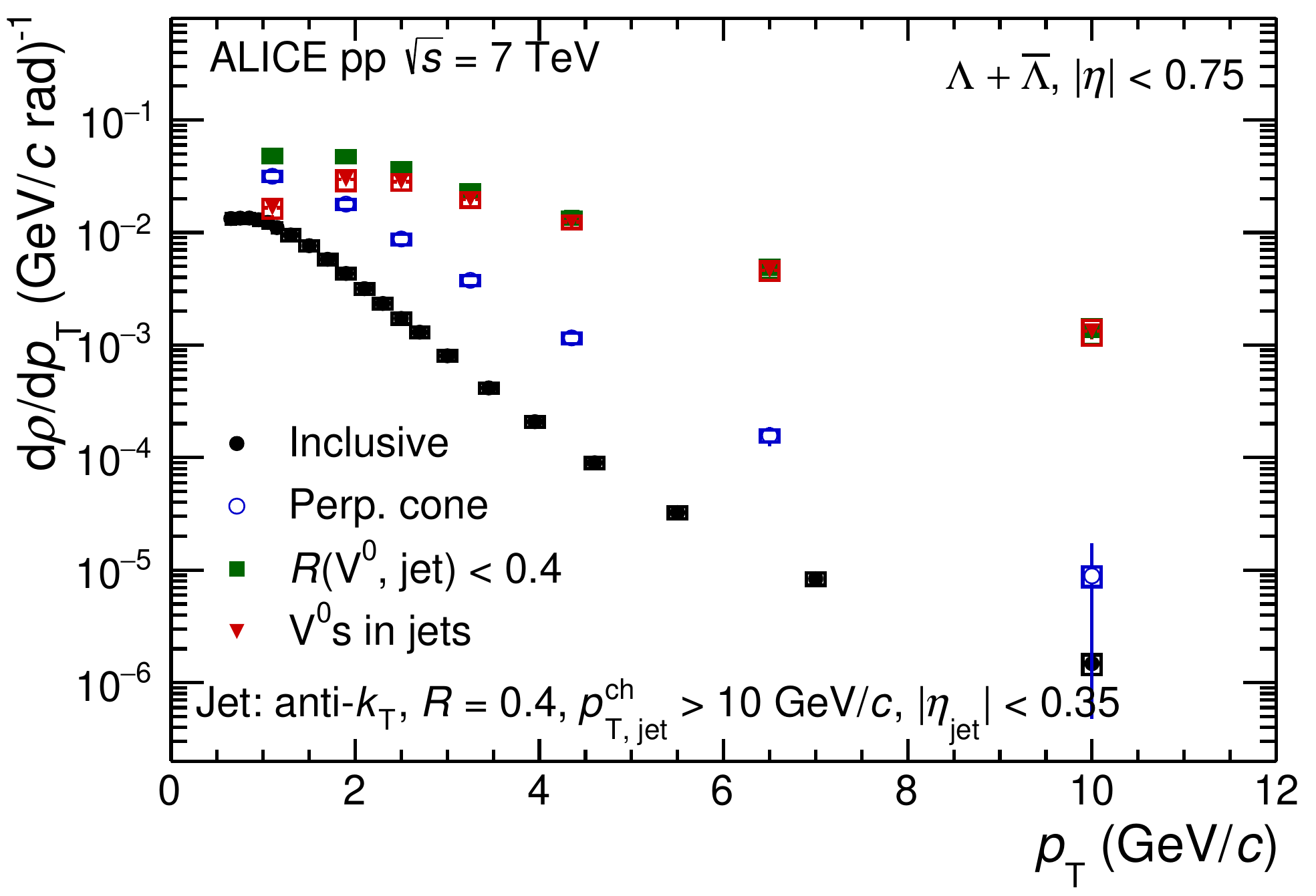}
\caption{The $\pt$-differential density of particles $\drhoVdpT$ (see Eq.~\eqref{eq:defv0rho}) in pp collisions at \seven for \kzero\ (upper), and the sum of \lmb\ and \almb\ (lower).
The density is shown for four selection criteria with the same definitions as Fig.~\ref{fig:rhov0pPb}.}
\label{fig:rhov0pp}
\end{figure}

The fully corrected densities of \kzero\ and the sum of \lmb\ and \almb\ particles associated with a hard scattering, tagged by a jet, are shown in Figs.~\ref{fig:rhov0pPb} and~\ref{fig:rhov0pp} for \pPb and pp collisions, respectively.
The per-jet density of \Vzero\ particles within jets is compared with that of inclusive particles (irrespective of their association with a hard scattering) and with underlying event \Vzero{}s obtained using the PC selection.
In the case of inclusive particles the distribution is normalised to the product of the total number of events and the acceptance of the \Vzero\ particles in a single event (full azimuth and $\abs{\hlab}<0.75$).
As expected, the $\pT$ dependence of the density of both \kzero\ and \lmb\ particles within jets, as defined by Eq.~\eqref{eq:defJEV0s}, is considerably less steep than in the case of inclusive particles.
The density distribution of inclusive $\Vzero$s is lower than that of the PC selection since the latter are obtained from events contain jets with $\pTjch > 10$~\GeVc.
But the density distribution of the PC selection shows a strong, steeply falling $\pT$ dependence with respect to the inclusive one.
Both the inclusive and the PC distributions show a rapid decrease with $\pT$, reaching values more than an order of magnitude lower than the JC density for particle $\pT$ exceeding $4$~\GeVc.
This is consistent with the expectation that the high-$\pT$ particles originate from jet fragmentation.

\begin{figure}[ht]
\centering
\includegraphics[width=0.80\textwidth]{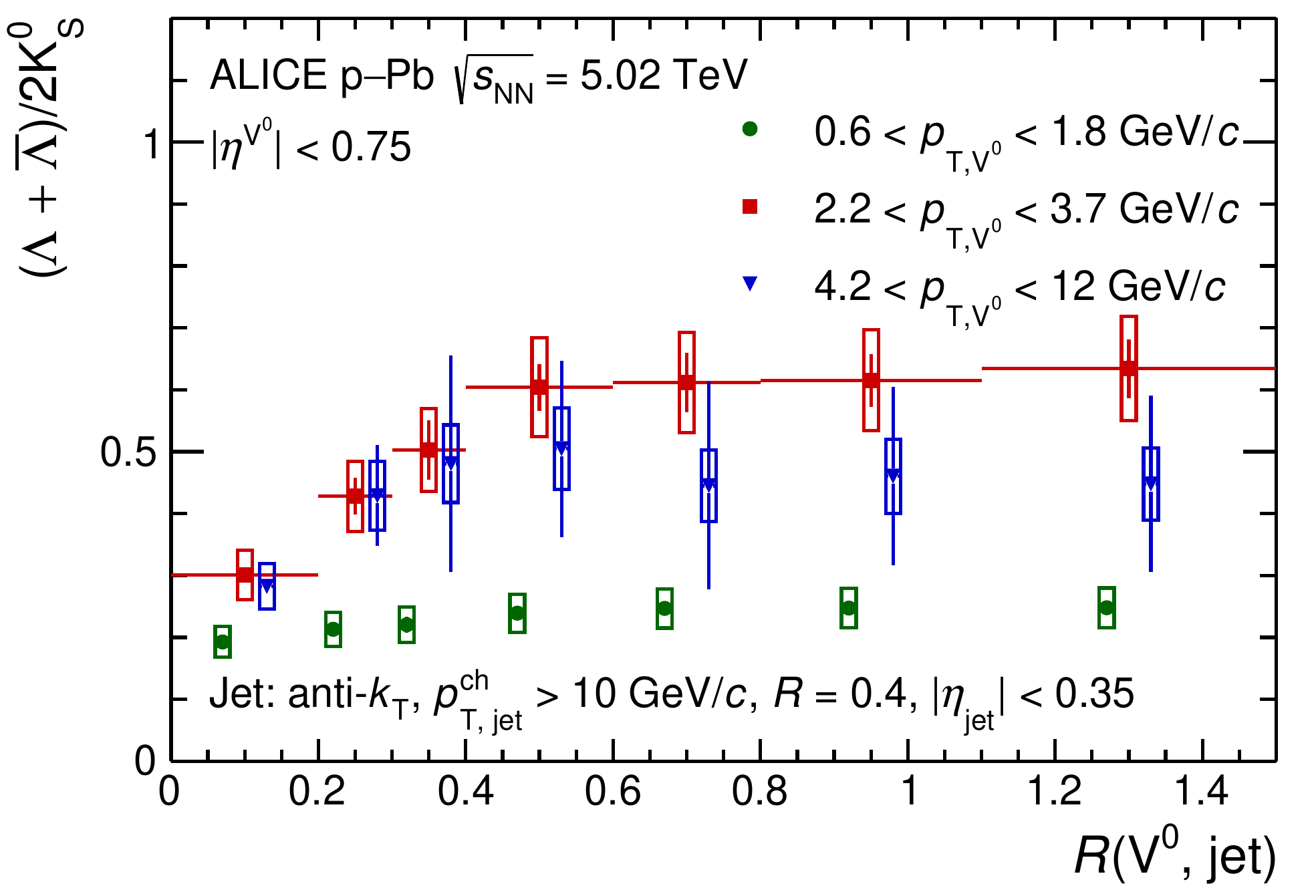}
\caption{The $\rLK$ ratio in \pPb collisions at \fivenn as a function of $\Rvj$ for three different \Vzero{}-particle $\pT$ intervals associated with charged jets with $\pTjch>10$~\GeVc.
The data points of the ratios in $0.6<\pTv <1.8$~\GeVc and in $4.2<\pTv <12$~\GeVc are shifted to the left and right sides from the centre, along the $\Rvj$-axis for better visibility.
Statistical uncertainties (vertical bars) and systematic uncertainties (open boxes) are shown.
The sources of the systematic uncertainty are summarized in table~\ref{tab:systR}.
The uncertainty on $\Vzero$ yield extraction is uncorrelated with $\Vzero$ $\pT$ but correlated with $\Rvj$, the uncertainties on jet $\pT$ scale and on UE subtraction are uncorrelated on both $\Vzero$ $\pT$ and $\Rvj$.}
\label{fig:LKR}
\end{figure}

Ratios of \lmb\ and \kzero\ yields can be obtained by dividing the normalised density distributions.
Here, the sum of the \lmb\ and \almb\ densities is divided by twice the density of \kzero.
Figure~\ref{fig:LKR} shows the ratio for the JC selection (without the UE background subtraction) as a function of the distance from the jet axis $\Rvj$ in \pPb collisions.
The ratio is shown for three $\pT$ intervals: low $\pT$ ($0.6 <\pT <1.8$~\GeVc), intermediate $\pT$ ($2.2 < \pT < 3.7$~\GeVc), and high $\pT$ ($4.2 < \pT < 12$~\GeVc).
The sources of the systematic uncertainties (open boxes) are summarized in Table~\ref{tab:systR}.
The uncertainty on $\Vzero$ yield extraction is uncorrelated with $\Vzero$ $\pT$ but correlated with $\Rvj$; the uncertainties on jet $\pT$ scale and on UE subtraction are uncorrelated on both $\Vzero$ $\pT$ and $\Rvj$.
The ratio as a function of $\Rvj$ at low $\pT$, dominated by the UE contribution, is approximately constant at about $0.2$.
It is independent of the distance to the jet axis even at large distances of $\Rvj > 1.2$.
This value is consistent with the inclusive measurements in \pPb\ collisions, but also in pp and peripheral \PbPb collisions where effects related to the collective expansion of the system are either not present or small~\cite{Acharya:2018orn}.

Conversely, the intermediate-$\pT$ selection shows an increase of the ratio from about $0.3$ when evaluated close to the jet axis to values of about $0.6$ at $\Rvj$ distances of about $0.5$.
For distances $\Rvj > 0.5$ the ratio remains constant.
The ratio of $0.6$ is consistent with the inclusive measurement in \pPb\ collisions~\cite{Abelev:2013haa} and this $\pT$ region is where the enhanced $\rLK$ ratio in the inclusive measurements is found to be the largest.
It is worthwhile to stress that for the results shown in Fig.~\ref{fig:LKR}, the UE backgrounds are not subtracted.
Therefore, the evolution of the ratio as a function of the distance from the jet axis demonstrates how the two sources, UE and jet, compete.
The lack of enhancement close to the jet axis indicates that the enhanced $\rLK$ ratio is not associated with jets.

\begin{figure}[ht]
\centering
\includegraphics[width=0.74\textwidth]{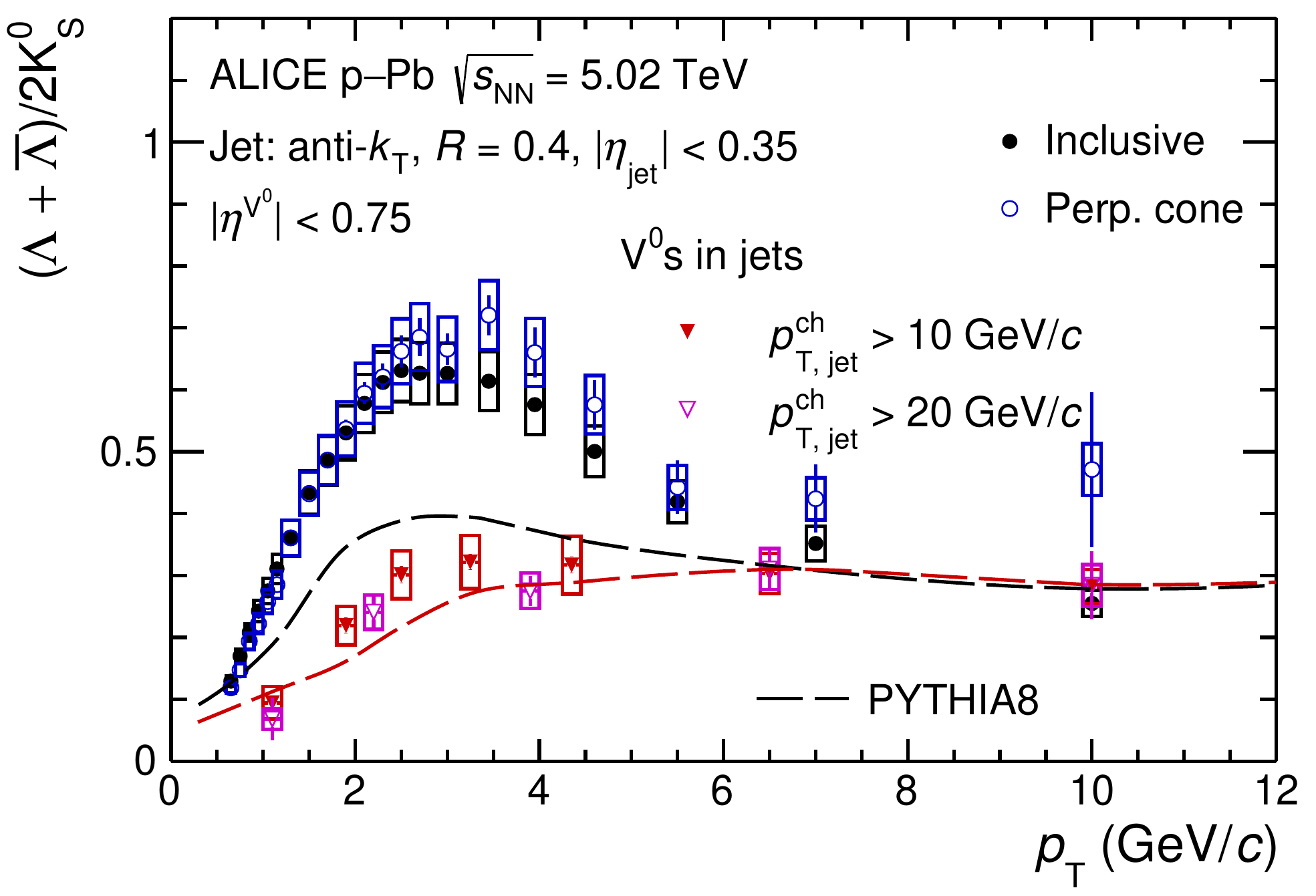} \\
\includegraphics[width=0.74\textwidth]{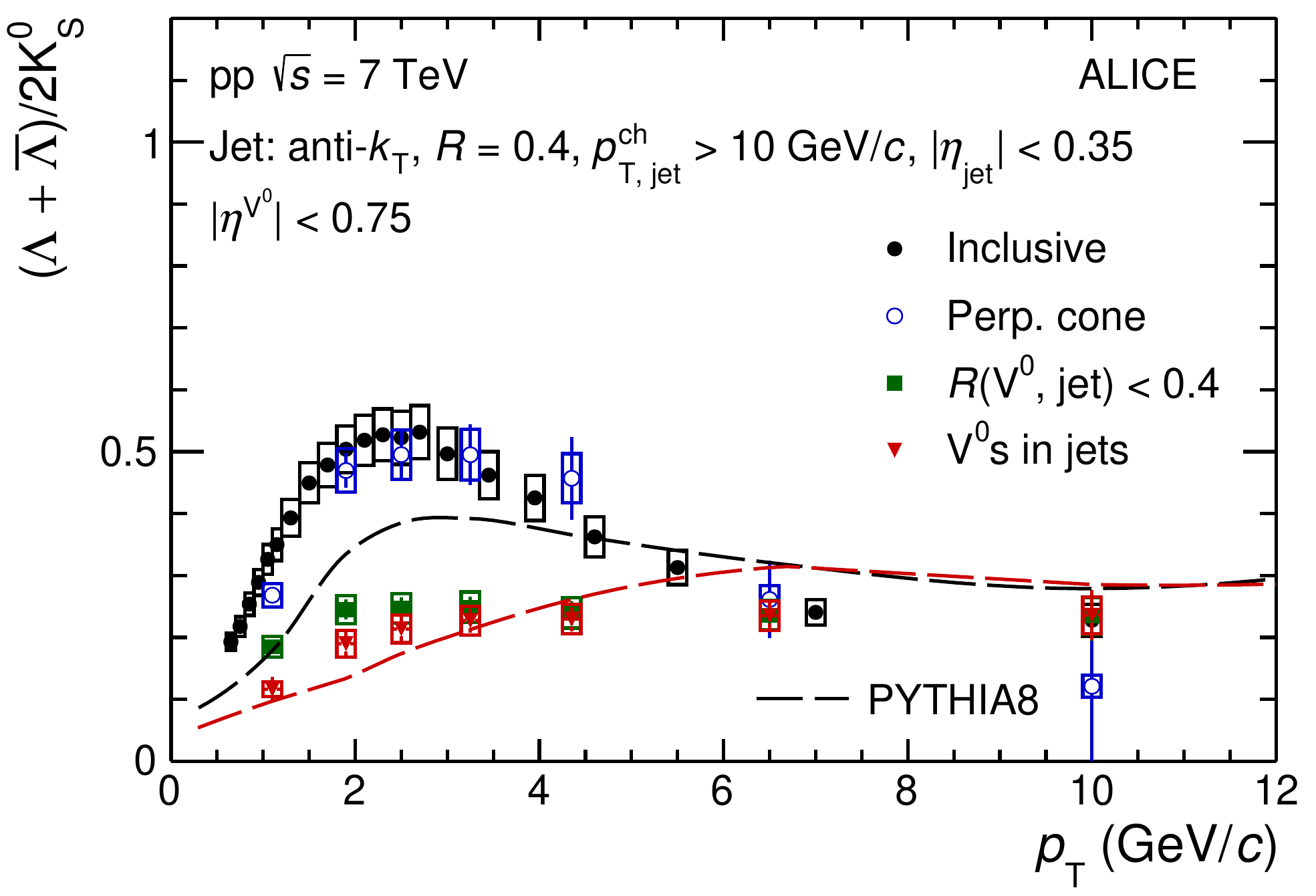}
\caption{The $\rLK$ ratio in \pPb collisions at \fivenn (upper panel) and pp collisions at \seven (lower panel) as a function of \Vzero{}-particle $\pT$, associated with charged jets with $\pTjch>10$~\GeVc (for both pp and p--Pb collisions) and $20$~\GeVc (for p--Pb collisions only) together with that in inclusive and PC selection, and JC selection in case of pp collisions.
The systematic uncertainties (open boxes) are fully uncorrelated with $\pT$.
In both upper and lower panels, the black dashed curves are the results for inclusive $\Vzero$s from \Pythia~$8$ simulations.
The jet selection within \Pythia~$8$ is made using the generator level information with $\pTjch>10$~\GeVc shown as the red curves.}
\label{fig:L2Kratio_pp_pPb}
\end{figure}

In each $\pT$ interval the ratio is dominated by the lower side of the selection window due to the steeply falling particle $\pT$ spectrum.
This is especially the case for $4.2 < \pT < 12$~\GeVc where the dominating component originates from $\pT$ of about $4.5$~\GeVc and the $\Rvj$ dependence at high $\pT$ is similar to that for $2.2 < \pT < 3.7$~\GeVc.
The ratio at high $\pT$ associated with jets is discussed below.

Figure~\ref{fig:L2Kratio_pp_pPb} shows the ratio of \lmb\ to \kzero\ as a function of particle $\pT$ in both pp and \pPb collisions for the different selection criteria.
The systematic uncertainties (open boxes) are fully uncorrelated with $\pT$.
In the case of \pPb\ collisions, the ratio of the inclusive particles, the particles from the PC selection, and for those within jet with resolution parameter $R = 0.4$ and $\pTjch>10$ and $>20$~\GeVc are shown.
Prior to forming the ratio, the UE density contribution obtained with the PC selection is subtracted for each particle species separately.
Additionally, the \pPb\ results are shown for the case where every \Vzero\ particle is required to be close to the jet axis with its distance $\Rvj < 0.4$.
The inclusive and the PC distributions show the enhancement at a $\pT$ of about $3$~\GeVc.
The measurement of the inclusive case differs from that in Ref.~\cite{Abelev:2013haa} as the region $\abs{\hlab}<0.75$ is used here instead of the rapidity region in centre-of-mass frame $0 < \ycms < 0.5$.
The two measurements are otherwise consistent with each other.
The PC distribution above $2$~\GeVc reaches systematically higher values than the inclusive.
The ratio within jets is consistently lower than the inclusive one and approximately independent of $\pT$ beyond $2$~\GeVc.
In particular, for particles associated with the jet it does not show a maximum at intermediate $\pT$.
Clearly the enhancement of the ratio seen in the inclusive measurement is not present within jets.
This conclusion holds not only for jets with $\pT>10$~\GeVc but also for higher $\pT$ ($>20$~\GeVc) jets.

The results for pp collisions shown in Fig.~\ref{fig:L2Kratio_pp_pPb} are obtained with jets reconstructed with $R=0.4$ and for the same value of the matching radius $\Rvj<0.4$.
Apart from the inclusive particle selection and UE selection, the figure shows the ratio for particles within jets for the UE subtracted in the JC and UE unsubtracted case, demonstrating the small magnitude of background effects.
Qualitatively similar features of the ratio are seen in both collision systems.

Selecting hard scatterings according to the jet energy carried exclusively by the primary charged particles induces biases and inefficiencies in the selection of the parton showers.
The bias is related to the probabilistic process of fragmentation and hadronization.
The analysis presented here tags only parton showers fragmenting into a configuration of hadrons that produce a charged particle jet with $\pTjch>10$~\GeVc with a given $R$ with a finite efficiency.
Therefore, there can be cases of \Vzero\ particles that originated from a parton shower but are rejected in the analysis based on the energy carried only by the primary charged particles.
The same analysis performed using the \Pythia~$8$ event generator shows that the most probable $\pT$ of the full jet with $R=0.4$ is larger by about 40\% as compared to the $\pTjch$.
Moreover, since the daughters of the \Vzero\ particles are not included in the jet energy calculation there are cases of jets containing \Vzero\ particles but not included in the JC selection.
On the other hand, Fig.~\ref{fig:L2Kratio_pp_pPb} shows that the inclusive $\rLK$ ratio at high $\pT$ is fully consistent with the ratio from particles associated with jets in this analysis. This suggests that the conclusion on the absence of the baryon-to-meson enhancement in jets made with the charged jets alone holds for all energetic parton showers and hadron configurations within jets.

\begin{figure}[!t]
\centering
\includegraphics[width=0.80\textwidth]{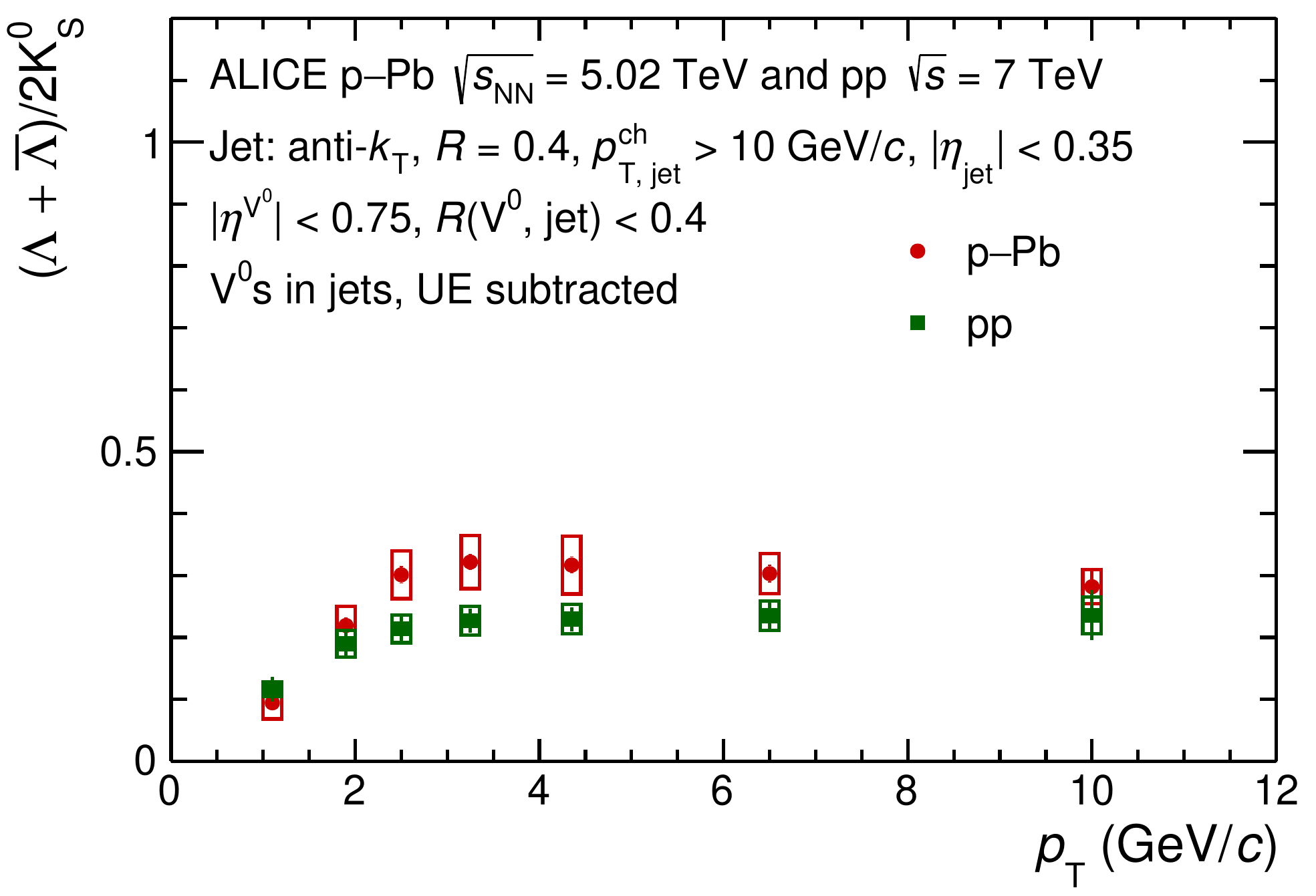}
\caption{The $\rLK$ ratio in pp collisions at \seven and in \pPb\  collisions at \fivenn as a function of \Vzero{}-particle $\pT$ associated with charged particle jets with $\pTjch>10$~\GeVc reconstructed using the \akT\ jet finder with resolution parameter $R=0.4$.
The ratio is shown for the same selection of the matching radius $\Rvj<0.4$ in both systems.
The systematic uncertainties (open boxes) are uncorrelated between the systems.}
\label{fig:L2Kratio_pp_pPb_comparison}
\end{figure}

Figure~\ref{fig:L2Kratio_pp_pPb} shows also the results compared with those obtained with the \Pythia~$8$~\cite{Sj_strand_2015} event generator with tune $4$C (the dashed curves) run for pp collisions at \five (top panel) and \seven (bottom panel).
The comparison shows that the characteristic maximum at intermediate $\pT$ in the inclusive ratio is not reproduced by the generator.
However, for both collision systems the ratio within jets after the subtraction of the underlying event is consistent with the data points within uncertainties for $\pT > 6$~\GeVc.
Note that \Pythia~$8$ was chosen here merely as an example and the aim is not for a thorough review of the strangeness production in the Monte Carlo generators.
The comparison with experimental data is found to be sufficient to demonstrate the clear similarities of the baryon-to-meson ratio within jets.

Figure~\ref{fig:L2Kratio_pp_pPb_comparison} shows the comparison of the ratio obtained in jets in pp and \pPb collisions for the same selection of the matching radius $\Rvj<0.4$ in both systems.
The ratio obtained in \pPb collisions is systematically higher for $2<\pT<8$~\GeVc with respect to that in pp collisions.
However, the difference between the two collision systems is less than $2\sigma$.
The deviation between pp and p--Pb collisions has to be studied with higher precision in the future.

\section{Summary}

The production of $\Vzero$ particles (\lmb\ baryons and \kzero\ mesons) is measured separately for particles associated with hard scatterings, tagged by reconstructed charged-particle jets, and the underlying event in \pPb collisions at \fivenn and pp collisions at \seven for the first time at the LHC.
The $\pT$-differential density distributions of $\Vzero$ particles associated within jets are compared with those obtained from inclusive analysis and the underlying event.
In both collision systems, the distribution of particles associated within jets is harder than that obtained in the underlying event since the high-$\pT$ particles originate from jet fragmentation.
The density of particles in the UE is larger than in the inclusive case as the former is obtained from events requiring a presence of a jet with $\pTjch > 10$~\GeVc.
The $\rLK$ ratio (without the UE subtracted) is studied as a function of $\Rvj$, defined as the distance between the jet axis and the $\Vzero$ particle, in \pPb collisions at \fivenn.
At intermediate $\pT$, the ratio increases with $\Rvj$ from a value about $0.3$ to $0.6$ up to $\Rvj = 0.5$ reaching a constant value of about $0.6$ for $\Rvj > 0.5$.
This demonstrates that the enhanced $\rLK$ ratio at intermediate $\pT$ observed in the inclusive analysis is not associated with jets since the underlying event contribution is more significant at larger $\Rvj$.
The $\rLK$ ratio associated with jets (with the UE subtracted) is consistent with the inclusive case within uncertainties for $\pT > 6$~\GeVc.
The results in p--Pb collisions for $\Rvj<0.4$ are consistent with the ratio measured in pp collisions.
Finally, the enhancement in the $\rLK$ ratio at intermediate $\pT$ found in the inclusive measurements in \pPb\ and \PbPb\ collisions is not present for particles associated with hard scatterings tagged by jets reconstructed from charged particles for $\pTjch>10$~\GeVc in \pPb\ and \pp\ collisions.
As the baryon-to-meson enhancement (``baryon anomaly'') found in the inclusive measurements has been linked to the interplay of radial flow and parton recombination at intermediate $\pt$, its absence within the jet cone demonstrates that these effects are indeed limited to the soft particle production processes.

\newenvironment{acknowledgement}{\relax}{\relax}
\begin{acknowledgement}
\section*{Acknowledgements}

The ALICE Collaboration would like to thank all its engineers and technicians for their invaluable contributions to the construction of the experiment and the CERN accelerator teams for the outstanding performance of the LHC complex.
The ALICE Collaboration gratefully acknowledges the resources and support provided by all Grid centres and the Worldwide LHC Computing Grid (WLCG) collaboration.
The ALICE Collaboration acknowledges the following funding agencies for their support in building and running the ALICE detector:
A. I. Alikhanyan National Science Laboratory (Yerevan Physics Institute) Foundation (ANSL), State Committee of Science and World Federation of Scientists (WFS), Armenia;
Austrian Academy of Sciences, Austrian Science Fund (FWF): [M 2467-N36] and Nationalstiftung f\"{u}r Forschung, Technologie und Entwicklung, Austria;
Ministry of Communications and High Technologies, National Nuclear Research Center, Azerbaijan;
Conselho Nacional de Desenvolvimento Cient\'{\i}fico e Tecnol\'{o}gico (CNPq), Financiadora de Estudos e Projetos (Finep), Funda\c{c}\~{a}o de Amparo \`{a} Pesquisa do Estado de S\~{a}o Paulo (FAPESP) and Universidade Federal do Rio Grande do Sul (UFRGS), Brazil;
Ministry of Education of China (MOEC) , Ministry of Science \& Technology of China (MSTC) and National Natural Science Foundation of China (NSFC), China;
Ministry of Science and Education and Croatian Science Foundation, Croatia;
Centro de Aplicaciones Tecnol\'{o}gicas y Desarrollo Nuclear (CEADEN), Cubaenerg\'{\i}a, Cuba;
Ministry of Education, Youth and Sports of the Czech Republic, Czech Republic;
The Danish Council for Independent Research | Natural Sciences, the VILLUM FONDEN and Danish National Research Foundation (DNRF), Denmark;
Helsinki Institute of Physics (HIP), Finland;
Commissariat \`{a} l'Energie Atomique (CEA) and Institut National de Physique Nucl\'{e}aire et de Physique des Particules (IN2P3) and Centre National de la Recherche Scientifique (CNRS), France;
Bundesministerium f\"{u}r Bildung und Forschung (BMBF) and GSI Helmholtzzentrum f\"{u}r Schwerionenforschung GmbH, Germany;
General Secretariat for Research and Technology, Ministry of Education, Research and Religions, Greece;
National Research, Development and Innovation Office, Hungary;
Department of Atomic Energy Government of India (DAE), Department of Science and Technology, Government of India (DST), University Grants Commission, Government of India (UGC) and Council of Scientific and Industrial Research (CSIR), India;
Indonesian Institute of Science, Indonesia;
Istituto Nazionale di Fisica Nucleare (INFN), Italy;
Institute for Innovative Science and Technology , Nagasaki Institute of Applied Science (IIST), Japanese Ministry of Education, Culture, Sports, Science and Technology (MEXT) and Japan Society for the Promotion of Science (JSPS) KAKENHI, Japan;
Consejo Nacional de Ciencia (CONACYT) y Tecnolog\'{i}a, through Fondo de Cooperaci\'{o}n Internacional en Ciencia y Tecnolog\'{i}a (FONCICYT) and Direcci\'{o}n General de Asuntos del Personal Academico (DGAPA), Mexico;
Nederlandse Organisatie voor Wetenschappelijk Onderzoek (NWO), Netherlands;
The Research Council of Norway, Norway;
Commission on Science and Technology for Sustainable Development in the South (COMSATS), Pakistan;
Pontificia Universidad Cat\'{o}lica del Per\'{u}, Peru;
Ministry of Education and Science, National Science Centre and WUT ID-UB, Poland;
Korea Institute of Science and Technology Information and National Research Foundation of Korea (NRF), Republic of Korea;
Ministry of Education and Scientific Research, Institute of Atomic Physics and Ministry of Research and Innovation and Institute of Atomic Physics, Romania;
Joint Institute for Nuclear Research (JINR), Ministry of Education and Science of the Russian Federation, National Research Centre Kurchatov Institute, Russian Science Foundation and Russian Foundation for Basic Research, Russia;
Ministry of Education, Science, Research and Sport of the Slovak Republic, Slovakia;
National Research Foundation of South Africa, South Africa;
Swedish Research Council (VR) and Knut \& Alice Wallenberg Foundation (KAW), Sweden;
European Organization for Nuclear Research, Switzerland;
Suranaree University of Technology (SUT), National Science and Technology Development Agency (NSDTA) and Office of the Higher Education Commission under NRU project of Thailand, Thailand;
Turkish Atomic Energy Agency (TAEK), Turkey;
National Academy of  Sciences of Ukraine, Ukraine;
Science and Technology Facilities Council (STFC), United Kingdom;
National Science Foundation of the United States of America (NSF) and United States Department of Energy, Office of Nuclear Physics (DOE NP), United States of America. 
\end{acknowledgement}

\bibliographystyle{utphys}
\bibliography{AliV0jets}

\newpage
\appendix

\section{The ALICE Collaboration}
\label{app:collab}

\begingroup
\small
\begin{flushleft}

S.~Acharya$^{\rm 142}$, 
D.~Adamov\'{a}$^{\rm 97}$, 
A.~Adler$^{\rm 75}$, 
J.~Adolfsson$^{\rm 82}$, 
G.~Aglieri Rinella$^{\rm 35}$, 
M.~Agnello$^{\rm 31}$, 
N.~Agrawal$^{\rm 55}$, 
Z.~Ahammed$^{\rm 142}$, 
S.~Ahmad$^{\rm 16}$, 
S.U.~Ahn$^{\rm 77}$
% ...

\section*{Affiliation notes}

$^{\rm I}$ Deceased\\
$^{\rm II}$ Also at: Italian National Agency for New Technologies, Energy and Sustainable Economic Development (ENEA), Bologna, Italy\\
$^{\rm IV}$ Also at: Dipartimento DET del Politecnico di Torino, Turin, Italy\\
$^{\rm VI}$ Also at: M.V. Lomonosov Moscow State University, D.V. Skobeltsyn Institute of Nuclear, Physics, Moscow, Russia\\
$^{\rm VII}$ Also at: Institute of Theoretical Physics, University of Wroclaw, Poland\\

\section*{Collaboration Institutes}

$^{1}$ A.I. Alikhanyan National Science Laboratory (Yerevan Physics Institute) Foundation, Yerevan, Armenia\\
$^{2}$ AGH University of Science and Technology, Cracow, Poland\\
$^{3}$ Bogolyubov Institute for Theoretical Physics, National Academy of Sciences of Ukraine, Kiev, Ukraine\\
$^{4}$ Bose Institute, Department of Physics  and Centre for Astroparticle Physics and Space Science (CAPSS), Kolkata, India\\
$^{5}$ Budker Institute for Nuclear Physics, Novosibirsk, Russia\\
$^{6}$ California Polytechnic State University, San Luis Obispo, California, United States\\
$^{7}$ Central China Normal University, Wuhan, China\\
$^{8}$ Centro de Aplicaciones Tecnol\'{o}gicas y Desarrollo Nuclear (CEADEN), Havana, Cuba\\
$^{9}$ Centro de Investigaci\'{o}n y de Estudios Avanzados (CINVESTAV), Mexico City and M\'{e}rida, Mexico\\
$^{10}$ Chicago State University, Chicago, Illinois, United States\\
$^{11}$ China Institute of Atomic Energy, Beijing, China\\
$^{12}$ Chungbuk National University, Cheongju, Republic of Korea\\
$^{13}$ Comenius University Bratislava, Faculty of Mathematics, Physics and Informatics, Bratislava, Slovakia\\
$^{14}$ COMSATS University Islamabad, Islamabad, Pakistan\\
$^{15}$ Creighton University, Omaha, Nebraska, United States\\
$^{16}$ Department of Physics, Aligarh Muslim University, Aligarh, India\\
$^{17}$ Department of Physics, Pusan National University, Pusan, Republic of Korea\\
$^{18}$ Department of Physics, Sejong University, Seoul, Republic of Korea\\
$^{19}$ Department of Physics, University of California, Berkeley, California, United States\\
$^{20}$ Department of Physics, University of Oslo, Oslo, Norway\\
$^{21}$ Department of Physics and Technology, University of Bergen, Bergen, Norway\\
$^{22}$ Dipartimento di Fisica dell'Universit\`{a} 'La Sapienza' and Sezione INFN, Rome, Italy\\
$^{23}$ Dipartimento di Fisica dell'Universit\`{a} and Sezione INFN, Cagliari, Italy\\
$^{24}$ Dipartimento di Fisica dell'Universit\`{a} and Sezione INFN, Trieste, Italy\\
$^{25}$ Dipartimento di Fisica dell'Universit\`{a} and Sezione INFN, Turin, Italy\\
$^{26}$ Dipartimento di Fisica e Astronomia dell'Universit\`{a} and Sezione INFN, Bologna, Italy\\
$^{27}$ Dipartimento di Fisica e Astronomia dell'Universit\`{a} and Sezione INFN, Catania, Italy\\
$^{28}$ Dipartimento di Fisica e Astronomia dell'Universit\`{a} and Sezione INFN, Padova, Italy\\
$^{29}$ Dipartimento di Fisica e Nucleare e Teorica, Universit\`{a} di Pavia  and Sezione INFN, Pavia, Italy\\
$^{30}$ Dipartimento di Fisica `E.R.~Caianiello' dell'Universit\`{a} and Gruppo Collegato INFN, Salerno, Italy\\
$^{31}$ Dipartimento DISAT del Politecnico and Sezione INFN, Turin, Italy\\
$^{32}$ Dipartimento di Scienze e Innovazione Tecnologica dell'Universit\`{a} del Piemonte Orientale and INFN Sezione di Torino, Alessandria, Italy\\
$^{33}$ Dipartimento di Scienze MIFT, Universit\`{a} di Messina, Messina, Italy\\
$^{34}$ Dipartimento Interateneo di Fisica `M.~Merlin' and Sezione INFN, Bari, Italy\\
$^{35}$ European Organization for Nuclear Research (CERN), Geneva, Switzerland\\
$^{36}$ Faculty of Electrical Engineering, Mechanical Engineering and Naval Architecture, University of Split, Split, Croatia\\
$^{37}$ Faculty of Engineering and Science, Western Norway University of Applied Sciences, Bergen, Norway\\
$^{38}$ Faculty of Nuclear Sciences and Physical Engineering, Czech Technical University in Prague, Prague, Czech Republic\\
$^{39}$ Faculty of Science, P.J.~\v{S}af\'{a}rik University, Ko\v{s}ice, Slovakia\\
$^{40}$ Frankfurt Institute for Advanced Studies, Johann Wolfgang Goethe-Universit\"{a}t Frankfurt, Frankfurt, Germany\\
$^{41}$ Fudan University, Shanghai, China\\
$^{42}$ Gangneung-Wonju National University, Gangneung, Republic of Korea\\
$^{43}$ Gauhati University, Department of Physics, Guwahati, India\\
$^{44}$ Helmholtz-Institut f\"{u}r Strahlen- und Kernphysik, Rheinische Friedrich-Wilhelms-Universit\"{a}t Bonn, Bonn, Germany\\
$^{45}$ Helsinki Institute of Physics (HIP), Helsinki, Finland\\
$^{46}$ High Energy Physics Group,  Universidad Aut\'{o}noma de Puebla, Puebla, Mexico\\
$^{47}$ Hiroshima University, Hiroshima, Japan\\
$^{48}$ Hochschule Worms, Zentrum  f\"{u}r Technologietransfer und Telekommunikation (ZTT), Worms, Germany\\
$^{49}$ Horia Hulubei National Institute of Physics and Nuclear Engineering, Bucharest, Romania\\
$^{50}$ Indian Institute of Technology Bombay (IIT), Mumbai, India\\
$^{51}$ Indian Institute of Technology Indore, Indore, India\\
$^{52}$ Indonesian Institute of Sciences, Jakarta, Indonesia\\
$^{53}$ INFN, Laboratori Nazionali di Frascati, Frascati, Italy\\
$^{54}$ INFN, Sezione di Bari, Bari, Italy\\
$^{55}$ INFN, Sezione di Bologna, Bologna, Italy\\
$^{56}$ INFN, Sezione di Cagliari, Cagliari, Italy\\
$^{57}$ INFN, Sezione di Catania, Catania, Italy\\
$^{58}$ INFN, Sezione di Padova, Padova, Italy\\
$^{59}$ INFN, Sezione di Roma, Rome, Italy\\
$^{60}$ INFN, Sezione di Torino, Turin, Italy\\
$^{61}$ INFN, Sezione di Trieste, Trieste, Italy\\
$^{62}$ Inha University, Incheon, Republic of Korea\\
$^{63}$ Institute for Gravitational and Subatomic Physics (GRASP), Utrecht University/Nikhef, Utrecht, Netherlands\\
$^{64}$ Institute for Nuclear Research, Academy of Sciences, Moscow, Russia\\
$^{65}$ Institute of Experimental Physics, Slovak Academy of Sciences, Ko\v{s}ice, Slovakia\\
$^{66}$ Institute of Physics, Homi Bhabha National Institute, Bhubaneswar, India\\
$^{67}$ Institute of Physics of the Czech Academy of Sciences, Prague, Czech Republic\\
$^{68}$ Institute of Space Science (ISS), Bucharest, Romania\\
$^{69}$ Institut f\"{u}r Kernphysik, Johann Wolfgang Goethe-Universit\"{a}t Frankfurt, Frankfurt, Germany\\
$^{70}$ Instituto de Ciencias Nucleares, Universidad Nacional Aut\'{o}noma de M\'{e}xico, Mexico City, Mexico\\
$^{71}$ Instituto de F\'{i}sica, Universidade Federal do Rio Grande do Sul (UFRGS), Porto Alegre, Brazil\\
$^{72}$ Instituto de F\'{\i}sica, Universidad Nacional Aut\'{o}noma de M\'{e}xico, Mexico City, Mexico\\
$^{73}$ iThemba LABS, National Research Foundation, Somerset West, South Africa\\
$^{74}$ Jeonbuk National University, Jeonju, Republic of Korea\\
$^{75}$ Johann-Wolfgang-Goethe Universit\"{a}t Frankfurt Institut f\"{u}r Informatik, Fachbereich Informatik und Mathematik, Frankfurt, Germany\\
$^{76}$ Joint Institute for Nuclear Research (JINR), Dubna, Russia\\
$^{77}$ Korea Institute of Science and Technology Information, Daejeon, Republic of Korea\\
$^{78}$ KTO Karatay University, Konya, Turkey\\
$^{79}$ Laboratoire de Physique des 2 Infinis, Ir\`{e}ne Joliot-Curie, Orsay, France\\
$^{80}$ Laboratoire de Physique Subatomique et de Cosmologie, Universit\'{e} Grenoble-Alpes, CNRS-IN2P3, Grenoble, France\\
$^{81}$ Lawrence Berkeley National Laboratory, Berkeley, California, United States\\
$^{82}$ Lund University Department of Physics, Division of Particle Physics, Lund, Sweden\\
$^{83}$ Moscow Institute for Physics and Technology, Moscow, Russia\\
$^{84}$ Nagasaki Institute of Applied Science, Nagasaki, Japan\\
$^{85}$ Nara Women{'}s University (NWU), Nara, Japan\\
$^{86}$ National and Kapodistrian University of Athens, School of Science, Department of Physics , Athens, Greece\\
$^{87}$ National Centre for Nuclear Research, Warsaw, Poland\\
$^{88}$ National Institute of Science Education and Research, Homi Bhabha National Institute, Jatni, India\\
$^{89}$ National Nuclear Research Center, Baku, Azerbaijan\\
$^{90}$ National Research Centre Kurchatov Institute, Moscow, Russia\\
$^{91}$ Niels Bohr Institute, University of Copenhagen, Copenhagen, Denmark\\
$^{92}$ Nikhef, National institute for subatomic physics, Amsterdam, Netherlands\\
$^{93}$ NRC Kurchatov Institute IHEP, Protvino, Russia\\
$^{94}$ NRC \guillemotleft Kurchatov\guillemotright  Institute - ITEP, Moscow, Russia\\
$^{95}$ NRNU Moscow Engineering Physics Institute, Moscow, Russia\\
$^{96}$ Nuclear Physics Group, STFC Daresbury Laboratory, Daresbury, United Kingdom\\
$^{97}$ Nuclear Physics Institute of the Czech Academy of Sciences, \v{R}e\v{z} u Prahy, Czech Republic\\
$^{98}$ Oak Ridge National Laboratory, Oak Ridge, Tennessee, United States\\
$^{99}$ Ohio State University, Columbus, Ohio, United States\\
$^{100}$ Petersburg Nuclear Physics Institute, Gatchina, Russia\\
$^{101}$ Physics department, Faculty of science, University of Zagreb, Zagreb, Croatia\\
$^{102}$ Physics Department, Panjab University, Chandigarh, India\\
$^{103}$ Physics Department, University of Jammu, Jammu, India\\
$^{104}$ Physics Department, University of Rajasthan, Jaipur, India\\
$^{105}$ Physikalisches Institut, Eberhard-Karls-Universit\"{a}t T\"{u}bingen, T\"{u}bingen, Germany\\
$^{106}$ Physikalisches Institut, Ruprecht-Karls-Universit\"{a}t Heidelberg, Heidelberg, Germany\\
$^{107}$ Physik Department, Technische Universit\"{a}t M\"{u}nchen, Munich, Germany\\
$^{108}$ Politecnico di Bari and Sezione INFN, Bari, Italy\\
$^{109}$ Research Division and ExtreMe Matter Institute EMMI, GSI Helmholtzzentrum f\"ur Schwerionenforschung GmbH, Darmstadt, Germany\\
$^{110}$ Russian Federal Nuclear Center (VNIIEF), Sarov, Russia\\
$^{111}$ Saha Institute of Nuclear Physics, Homi Bhabha National Institute, Kolkata, India\\
$^{112}$ School of Physics and Astronomy, University of Birmingham, Birmingham, United Kingdom\\
$^{113}$ Secci\'{o}n F\'{\i}sica, Departamento de Ciencias, Pontificia Universidad Cat\'{o}lica del Per\'{u}, Lima, Peru\\
$^{114}$ St. Petersburg State University, St. Petersburg, Russia\\
$^{115}$ Stefan Meyer Institut f\"{u}r Subatomare Physik (SMI), Vienna, Austria\\
$^{116}$ SUBATECH, IMT Atlantique, Universit\'{e} de Nantes, CNRS-IN2P3, Nantes, France\\
$^{117}$ Suranaree University of Technology, Nakhon Ratchasima, Thailand\\
$^{118}$ Technical University of Ko\v{s}ice, Ko\v{s}ice, Slovakia\\
$^{119}$ The Henryk Niewodniczanski Institute of Nuclear Physics, Polish Academy of Sciences, Cracow, Poland\\
$^{120}$ The University of Texas at Austin, Austin, Texas, United States\\
$^{121}$ Universidad Aut\'{o}noma de Sinaloa, Culiac\'{a}n, Mexico\\
$^{122}$ Universidade de S\~{a}o Paulo (USP), S\~{a}o Paulo, Brazil\\
$^{123}$ Universidade Estadual de Campinas (UNICAMP), Campinas, Brazil\\
$^{124}$ Universidade Federal do ABC, Santo Andre, Brazil\\
$^{125}$ University of Cape Town, Cape Town, South Africa\\
$^{126}$ University of Houston, Houston, Texas, United States\\
$^{127}$ University of Jyv\"{a}skyl\"{a}, Jyv\"{a}skyl\"{a}, Finland\\
$^{128}$ University of Kansas, Lawrence, Kansas, United States\\
$^{129}$ University of Liverpool, Liverpool, United Kingdom\\
$^{130}$ University of Science and Technology of China, Hefei, China\\
$^{131}$ University of South-Eastern Norway, Tonsberg, Norway\\
$^{132}$ University of Tennessee, Knoxville, Tennessee, United States\\
$^{133}$ University of the Witwatersrand, Johannesburg, South Africa\\
$^{134}$ University of Tokyo, Tokyo, Japan\\
$^{135}$ University of Tsukuba, Tsukuba, Japan\\
$^{136}$ Universit\'{e} Clermont Auvergne, CNRS/IN2P3, LPC, Clermont-Ferrand, France\\
$^{137}$ Universit\'{e} de Lyon, CNRS/IN2P3, Institut de Physique des 2 Infinis de Lyon , Lyon, France\\
$^{138}$ Universit\'{e} de Strasbourg, CNRS, IPHC UMR 7178, F-67000 Strasbourg, France, Strasbourg, France\\
$^{139}$ Universit\'{e} Paris-Saclay Centre d'Etudes de Saclay (CEA), IRFU, D\'{e}partment de Physique Nucl\'{e}aire (DPhN), Saclay, France\\
$^{140}$ Universit\`{a} degli Studi di Foggia, Foggia, Italy\\
$^{141}$ Universit\`{a} di Brescia and Sezione INFN, Brescia, Italy\\
$^{142}$ Variable Energy Cyclotron Centre, Homi Bhabha National Institute, Kolkata, India\\
$^{143}$ Warsaw University of Technology, Warsaw, Poland\\
$^{144}$ Wayne State University, Detroit, Michigan, United States\\
$^{145}$ Westf\"{a}lische Wilhelms-Universit\"{a}t M\"{u}nster, Institut f\"{u}r Kernphysik, M\"{u}nster, Germany\\
$^{146}$ Wigner Research Centre for Physics, Budapest, Hungary\\
$^{147}$ Yale University, New Haven, Connecticut, United States\\
$^{148}$ Yonsei University, Seoul, Republic of Korea\\

\bigskip 

\end{flushleft} 
\endgroup 

\end{document}